\RequirePackage{ifpdf}
\documentclass[letterpaper]{JHEP3}
\usepackage{amsmath}
\usepackage{epsfig}
\usepackage{subfigure}

\DeclareSymbolFont{matha}{OML}{txmi}{m}{it}
\DeclareMathSymbol{\varv}{\mathord}{matha}{118}

\def\wc{y}
\def\wth{\vartheta}
\def\wlmb{\alpha_0}
\def\wvph{\theta}
\def\wcF{\check A}
\def\wQ{Q}

\newcommand{\roughly}[1]{\mathrel{\raise.3ex\hbox{$#1$\kern-0.85em
\lower1ex\hbox{$\sim$}}}}

\newcommand{\lsim}{\roughly<}
\newcommand{\gsim}{\roughly>}
\def\pd{\partial}

\def\la{{\langle}}
\def\ra{{\rangle}}

\def\cC{{\cal C}}

\def\cX{{\cal X}}

\def\cG{{\cal G}}

\def\cL{{\cal L}}

\def\cN{{\cal N}}
\def\cO{{\cal O}}
\def\cP{{\cal P}}
\def\cR{{\cal R}}
\def\cT{{\cal T}}

\def\cV{{\cal V}}

\def\cZ{{\cal Z}}

\newbox\charbox
\newbox\slabox
\def\slsh#1{{      % Feynman slash
        \setbox\charbox=\hbox{$#1$}
        \setbox\slabox=\hbox{$/$}
        \dimen\charbox=\ht\slabox
        \advance\dimen\charbox by -\dp\slabox
        \advance\dimen\charbox by -\ht\charbox
        \advance\dimen\charbox by \dp\charbox
        \divide\dimen\charbox by 2
        \raise-\dimen\charbox\hbox to \wd\charbox{\hss/\hss}
        \llap{$#1$}
}}

\def\exd{{\hbox{d}}}
\def\d{\exd}

\def\bea{\begin{eqnarray}}
\def\eea{\end{eqnarray}}
\def\be{\begin{equation}}
\def\ee{\end{equation}}

\def\ssA{{\scriptscriptstyle A}}
\def\ssB{{\scriptscriptstyle B}}

\def\ssD{{\scriptscriptstyle D}}
\def\ssE{{\scriptscriptstyle E}}

\def\ssH{{\scriptscriptstyle H}}

\def\ssM{{\scriptscriptstyle M}}
\def\ssN{{\scriptscriptstyle N}}
\def\ssP{{\scriptscriptstyle P}}

\def\ssV{{\scriptscriptstyle V}}

\def\ssZ{{\scriptscriptstyle Z}}
\def\BLF{{\scriptscriptstyle BLF}}

\def\EH{{\scriptscriptstyle EH}}
\def\EM{{\scriptscriptstyle EM}}

\def\nn{\nonumber}

\def\d{\mathrm{d}}

\def\({\left(}
\def\){\right)}

\def\pref#1{(\ref{#1})}

\def\d{\mathrm{d}}

\numberwithin{equation}{section}

\title{The Gravity of Dark Vortices: Effective Field Theory for Branes and Strings Carrying Localized Flux}

\author{C.P.~Burgess,$^{1,2,3}$ R.~Diener${}^{1,2}$ and M. Williams$^{4}$ \\
$^1$ Physics \& Astronomy, McMaster University,
Hamilton, ON, Canada, L8S 4M1\\
$^2$ Perimeter Institute for Theoretical Physics, Waterloo, ON, Canada N2L 2Y5\\
${}^3$ Division PH\,-TH, CERN, CH-1211, Gen\`eve 23, Suisse\\
$^4$ Instituut voor Theoretische Fysica, KU Leuven,
B-3001 Leuven, Belgium
}

\preprint{Preprint:CERN-PH-TH-2015-053}
\date{\today}

\abstract { A Nielsen-Olesen vortex usually sits in an environment that expels the flux that is confined to the vortex, so flux is not present both inside and outside. We construct vortices for which this is not true, where the flux carried by the vortex also permeates the `bulk' far from the vortex. The idea is to mix the vortex's internal gauge flux with an external flux using off-diagonal kinetic mixing. Such `dark' vortices could play a phenomenological role in models with both cosmic strings and a dark gauge sector. When coupled to gravity they also provide explicit ultra-violet completions for codimension-two brane-localized flux, which arises in extra-dimensional models when the same flux that stabilizes extra-dimensional size is also localized on space-filling branes situated around the extra dimensions. We derive simple formulae for observables such as defect angle, tension, localized flux and on-vortex curvature when coupled to gravity, and show how all of these are insensitive to much of the microscopic details of the solutions, and are instead largely dictated by low-energy quantities. We derive the required effective description in terms of a world-sheet brane action, and derive the matching conditions for its couplings. We consider the case where the dimensions transverse to the bulk compactify, and determine how the on- and off-vortex curvatures and other bulk features depend on the vortex properties. We find that the brane-localized flux does not gravitate, but just renormalizes the tension in a magnetic-field independent way. The existence of an explicit UV completion puts the effective description of these models on a more precise footing, verifying that brane-localized flux can be consistent with sensible UV physics and resolving some apparent paradoxes that can arise with a naive (but commonly used) delta-function treatment of the brane's localization within the bulk. }

\begin{document}

\section{Introduction}

In this paper we study the gravitational response of vortices that carry localized amounts of external magnetic flux; called {\em Dark Strings} or {\em Dark Vortices} in the literature \cite{DStrings, DSGravity}. The goal is to understand how their back-reaction influences the transverse geometry through which they move, and the geometry that is induced on their own world-sheet. We find the initially surprising result that the gravitational response of such an object is locally independent of the amount of flux it contains, and show how this can be simply understood.

\subsubsection*{Motivation}

Why study the gravitational response of Dark Vortices?

Vortices are among the simplest stable solitons and arise in many theories with spontaneously broken $U(1)$ gauge symmetries \cite{NOSolns}. They can arise cosmologically as relics of epochs when the Universe passes through symmetry-breaking phase transitions. Such cosmic strings are widely studied \cite{CStrings} because, unlike other types of cosmic defects, they need not be poisonous for later cosmology since the resulting cosmic tangle tends not to come to dominate the energy density in a problematic way.

In the simplest models a vortex defines a region outside of which the $U(1)$ symmetry breaks while inside it remains (relatively) unbroken, and as a result all magnetic $U(1)$ flux is confined to lie completely within the vortex interior. However in theories with more than one $U(1)$ factor more complicated patterns can also exist, for which magnetic fields outside the vortex can also acquire a localized intra-vortex component. Such vortices naturally arise in `Dark Photon' models \cite{DPhoton}, for which the ordinary photon mixes kinetically \cite{Bob} with a second, spontaneously broken, $U(1)$ gauge field (as have been widely studied as Dark Matter candidates \cite{DPDarkMatter}). Cosmic strings of this type could carry localized ordinary magnetic flux, even though the $U(1)_\EM$ gauge group remains unbroken \cite{DStrings, DSGravity}.

Of most interest are parameters where the vortex's transverse thickness is much smaller than the sizes of interest for the geometry transverse to the source. In such situations only a few vortex properties are important, including the tension (energy per unit length) and the amount of flux localized on the vortex (or more generally brane-localized flux, or BLF for short). Indeed these two quantities (call them $T_b$ and $\zeta_b$) provide the coefficients of the leading terms in any derivative expansion of a vortex action (for which more explicit forms are also given below),
\be \label{Sbforms}
 S_b = - T_b \, \int  \omega +  \zeta_b \int \, \star \, A + \cdots \,,
\ee
where $\omega$ is the volume form of the codimension-two surface and $\star A$ is the Hodge dual of the $U(1)$ field strength, $A_{\ssM \ssN} = \partial_\ssM A_\ssN - \partial_\ssN A_\ssM$ whose flux is carried by the vortex. These are the leading terms inasmuch as all terms represented by the ellipses involve two or more derivatives.\footnote{A single-derivative term involving the world-sheet extrinsic curvature is also possible, but our focus here is on straight motionless vortices.} In four dimensions both $\omega$ and $\star A$ are 2-forms and so can be covariantly integrated over the 2-dimensional world-sheet of a cosmic string, while in $D=d+2$ dimensions they are $d$ forms that can be integrated over the $d$-dimensional world volume of a codimension-2 surface.\footnote{That is, a brane with precisely two transverse off-brane dimensions.} Previous workers have studied gravitational response in the absence of brane-localized flux \cite{OtherGVs}, but our particular interest is on how $\zeta_b$ competes with $T_b$ to influence the geometry. Our analysis extends recent numerical studies \cite{DSGravity} of how dark strings gravitate, including in particular an effective field theory analysis of the BLF term and its gravitational properties.

Besides being of practical interest for Dark Photon models, part of our motivation for this study also comes from brane-world models within which the familiar particles of the Standard model reside on a 3+1 dimensional brane or `vortex' within a higher-dimensional space.\footnote{Our restriction to codimension-2 branes makes $d=4$ and $D=6$ the most interesting case of this type \cite{GS}.} Comparatively little is known about how higher-codimension branes situated within compact extra dimensions back-react gravitationally to influence their surrounding geometries,\footnote{By contrast, back-reaction is fairly well-explored for codimension-1 objects due to the extensive study of Randall-Sundrum models \cite{RS}.} and codimension-2 objects provide a simple nontrivial starting point for doing so. In particular, a key question in any such model is what stabilizes the size and shape of the transverse compact dimensions, and this is a question whose understanding can hinge on understanding how the geometry responds to the presence of the branes. Since long-range inter-brane forces vary only logarithmically in two transverse dimensions, they do not fall off with distance and so brane back-reaction and inter-brane forces are comparatively more important for codimension-2 objects than they are with more codimensions.

Furthermore, several mechanisms are known for stabilizing extra dimensions, and the main ones involve balancing inter-brane gravitational forces against the cost of distorting extra-dimensional cycles wrapped by branes or threaded by topological fluxes \cite{SS, GKP, GoldWis, 6DStabnb, 6DStab}. Since brane-localized flux is the leading way fluxes and uncharged branes directly couple to one another, it is crucial for understanding how flux-carrying vortices interact with one another and their transverse environment. Localized flux has recently also been recognized to play a role in the stability of compact geometries \cite{dahlen}.

Finally, the fact that cosmic strings can have flat world-sheets for any value of their string tension \cite{OtherGVs} has been used to suggest \cite{CLP, CG} they may contain the seeds of a mechanism for understanding the cosmological constant problem \cite{CCprob}. But a solution to the cosmological constant problem involves also understanding how the curvature of the world-sheet varies as its tension and other properties vary. This requires a critical study of how codimension-2 objects back-react onto their own induced geometry, such as we give here. Although extra-dimensional branes are not in themselves expected to be sufficient to provide a solution (for instance, one must also deal with the higher-dimensional cosmological constant), the techniques developed here can also be applied to their supersymmetric alternatives \cite{SLED}, for which higher-derivative cosmological constants are forbidden by supersymmetry and whose ultimate prospects remain open at this point. We make this application in a companion paper \cite{Companion}.

\subsubsection*{Results}

Our study leads to the following result:~{\em brane-localized flux does not gravitate.} It is most intuitively understood when it is the dual field $F = \star\, A$ that is held fixed when varying the metric, since in this case the BLF term $S_\BLF = \zeta \int F$ is metric-independent. We show how the same result can also be seen when $A$ is fixed; and more precisely show that the $\zeta_b$ (or BLF) term of \pref{Sbforms} induces a universal renormalization of the brane's tension and the brane gravitational response is governed only by the total tension including this renormalization. This renormalization is universal in the sense that it does not depend on the size of any macroscopic magnetic field in which the vortex may sit. (The central discussion, with equations, can be found between eqs.~\pref{Tzeta-UV} and \pref{endofhighlight}, below.)

Of course the BLF term {\em does} contribute to the external Maxwell equations, generating a flux localized at the vortex position with size proportional to $\zeta_b$. Among other things this ensures that a test charge that moves around the vortex acquires the Aharonov-Bohm phase implied by the localized flux. But its gravitational influence is precisely cancelled by the back-reaction of the Maxwell field, through the gravitational field set up by the localized flux to which the BLF term gives rise. Since an external macroscopic observer cannot resolve the energy of the vortex-localized BLF term from the energy of the localized magnetic field to which it gives rise, macroscopic external gravitational measurements only see their sum, which is zero.

The presence of the localized energy in the induced magnetic field does change the total energy density of the vortex, however, which can be regarded as renormalizing the vortex tension. This renormalization is independent of the strength of any outside magnetic fields.

This failure of the BLF term to gravitate has important implications for the curvature that is induced on the vortex world-sheet. To see why, consider the trace-reversed Einstein equations in $D = d+2$ dimensions, which state\footnote{We use Weinberg's curvature conventions \cite{Wbg}, which differ from those of MTW \cite{MTW} only by an overall sign in the definition of the Riemann tensor. Coordinates $x^\ssM$ label all $D$ dimensions while $x^\mu$ ($x^m$) label the $d$-dimensional (2-dimensional) subspaces.}
\be
 R_{\ssM\ssN} + \kappa^2 \left( T_{\ssM\ssN} - \frac{1}{d} \, g_{\ssM\ssN} \, {T^\ssP}_\ssP \right) = 0 \,.
\ee
What is special about this equation is that the factor of $1/d$ ensures that the on-brane stress-energy often drops out of the expression for the on-brane curvature, which is instead governed purely by the {\em off}-brane stress energy. Consequently it is of particular interest to know when $T_{mn}$ vanishes for some reasonable choice of brane lagrangian.

$T_{mn}$ would vanish in particular when the brane action is dominated by its tension
\be \label{tensionSE}
 T_{\mu\nu} = T_b \, g_{\mu\nu} \; \frac{\delta(y)}{\sqrt{g_2}} \,,
\ee
where $\delta(y)$ is some sort of regularized delta-like function with support only at the brane position. But the derivation of \pref{tensionSE} from \pref{Sbforms} is complicated by two issues: is there a dependence on the transverse metric hidden in the regularized $\delta(y)$ (which is designed, after all, to discriminate based on proper distance from the vortex); and (for flux-containing branes) what of the metrics appearing in the Hodge dual, $\star A$, of the BLF term?

The results found here imply these two issues are not obstructions to deriving \pref{tensionSE} from \pref{Sbforms}. They do this in two ways. First they show how $T_{mn}$ can be derived without ad-hoc assumptions about the metric-dependence of $\delta(y)$. Second, they show that the apparent dependence of the BLF terms on the transverse metric components, $g_{mn}$, is an illusion, because it is completely cancelled by a similar dependence in the gauge-field back-reaction.

The remainder of this paper shows how this works in detail. We use three different techniques to do so.
\begin{itemize}
\item The first works within a UV completion of the dark vortex, for which we explicitly solve all field equations for a system that allows Nielsen-Olesen type vortex solutions. In this construction the BLF term can arise if there is a kinetic mixing, $\varepsilon Z_{\ssM \ssN} A^{\ssM \ssN}$, between the $U(1)$ gauge field, $Z_\ssM$, of the Nielsen-Olesen vortex, and the external gauge field, $A_\ssM$, whose flux is to be localized. In this case the mixing of the two gauge fields can be diagonalized explicitly, leading to the advertised cancellation of the BLF coupling as well as a renormalization of the $Z_\ssM$ gauge coupling, $e^2 \to \hat e^2 = e^2 / (1 - \varepsilon^2)$.
\item Second, we compute the couplings $T$ and $\zeta$ of the effective action for the codimension-2 vortex in the limit where the length scales of the transverse geometry are much larger than the vortex size. This has the form of \pref{Sbforms}, with $\zeta_b \propto \varepsilon/e$. We verify that it reproduces the physics of the full UV theory, including in particular the cancellation of BLF gravitational interaction and the renormalization of the brane tension quadratically in $\zeta$.
\item Finally we compare both of these approaches to explicit numerical calculations of the metric-vortex profiles as functions of the various external parameters, to test the robustness of our results.
\end{itemize}

\subsubsection*{A road map}

The remainder of the paper is organized as follows.

The next section, \S\ref{section:system}, describes the action and field equations for the microscopic (or UV) system of interest. \S\ref{subsec:actionFE} shows this consists of a `bulk' sector (the metric plus a gauge field, $A_\ssM$) coupled to a `vortex' sector (a charged scalar, $\Psi$, and a second gauge field, $Z_\ssM$). The vortex sector is designed to support Nielsen-Olesen vortices and these provide the microscopic picture of how the codimension-2 objects arise. The symmetry ans\"atze used for these solutions are described in \S\ref{subsec:ansatz} and the order-of-magnitude scales given by the parameters of the system are summarized in \S\ref{subsec:scales}.

Solutions to the field equations describing a single isolated vortex are then described in detail in \S\ref{section:isolatedvortex}, including both analytic and numerical results for the field profiles. The logic of this section, starting in \S\ref{subsec:vortexsoln}, is to integrate the field equations in the radial direction, starting from initial conditions at the centre of the vortex and working our way out. The goal is to compute the values of the fields and their first derivatives just outside the vortex. In general we find a three-parameter set of choices for initial conditions (modulo coordinate conditions), that can be taken to be the flux quantum, $n$, for the vortex together with two integration constants ($Q$ and $\check R$) that describe the size of the ambient external magnetic field and the curvature of the on-brane directions.\footnote{For a given vortex lagrangian the tension of the vortex is controlled in terms of $n$ by parameters in the lagrangian. We can also take the tension to be a separate dial -- independent of $n$ --- if we imagine having several vortex sectors with different coupling constants in each sector.} The resulting formulae for the fields and derivatives external to the vortex provide the initial data for further integration into the bulk, and are efficiently captured through their implications for the asymptotic near-vortex form of the bulk solutions, described in \S\ref{subsec:nearvortex}. In \S\ref{subsec:vortexEFT} these expressions for the near-vortex fields and derivatives are also used to match with the effective vortex description of \pref{Sbforms} to infer expressions for $T_b$ and $\zeta_b$ in terms of microscopic parameters.

The point of view shifts in \S\ref{section:interactions} from the perspective of a single vortex to the question of how the bulk responds once the two vortices at each end are specified.\footnote{Using electrostatics in 3 spatial dimensions as an analogy, \S\ref{section:isolatedvortex} does the analog of relating the coefficient of $1/r$ in the electrostatic potential to the charge defined by the properties of the source. Then \S\ref{section:interactions} asks what the equilibrium configuration is for a collection of charges given the resulting electrostatic potential.} This is done in two ways. One can either continue integrating the field equations radially away from the first source (with $n$, $Q$ and $\check R$ specified as initial data as before) and thereby learn the properties of the source at the other end of the transverse space (by studying the singularities of the geometry where it closes off and compactifies). Alternatively, one can take the properties of the two sources as given and instead infer the values of $Q$ and $\check  R$ that are consistent with the source properties: the two flux quanta $n_+$ and $n_-$, and the overall quantum $N$ for the total magnetic flux in the transverse dimensions. After \S\ref{subsec:integrals} first provides a set of exact integral expressions for quantities like $\check  R$ in terms of other properties of the source and bulk solutions, \S\ref{subsec:Exactsolns} describes the exact solutions for the bulk that are maximally symmetric in the on-brane directions and interpolate between any pair of source vortices.

Finally, \S\ref{section:discussion} summarizes our results and describes several open directions. Some useful but subsidiary details of the calculations are given in several Appendices.

\section{The system of interest}
\label{section:system}

We start by outlining the action and field equations for the system of interest. Our system consists of an Einstein-Maxwell system (the `bulk') coupled to a `vortex' --- or `brane' --- sector, consisting of a complex scalar coupled to a second $U(1)$ gauge field. For generality we imagine both of these systems live in $D = d+2$ spacetime dimensions, though the most interesting cases of practical interest are the cosmic string [with $(D,d) = (4,2)$] and the brane-world picture [with $(D,d) = (6,4)$].

\subsection{Action and field equations}
\label{subsec:actionFE}

The action of interest is $S = S_\ssB + S_\ssV$ with bulk action given by
\bea \label{SB}
 S_\ssB &=& - \int \exd^{d+2}x \; \sqrt{-g} \left[ \frac{1}{2\kappa^2} \; g^{\ssM\ssN} \, \cR_{\ssM \ssN} + \frac14 \, A_{\ssM \ssN} A^{\ssM \ssN} + \Lambda \right] \nn\\
 &=:& - \int \exd^{d+2}x \; \sqrt{-g} \; \Bigl( L_\EH + L_\ssA + \Lambda \Bigr)
\eea
where $A_{\ssM \ssN} = \partial_\ssM A_\ssN - \partial_\ssN A_\ssM$ is a $D$-dimensional gauge field strength, $\cR_{\ssM\ssN}$ denotes the $D$-dimensional Ricci tensor and the last line defines the $L_i$ in terms of the corresponding item in the previous line. The vortex part of the action is similarly given by
\bea \label{SV}
 S_\ssV &=& - \int \exd^{d+2}x \; \sqrt{-g} \left[ \frac14 \, Z_{\ssM \ssN} Z^{\ssM \ssN} + \frac{\varepsilon}2 \, Z_{\ssM \ssN} A^{\ssM \ssN} + D_\ssM \Psi^* \, D^\ssM \Psi  + \lambda \, \left(\Psi^* \Psi - \frac{v^2}{2} \right)^2 \right] \nn\\
 &=:& - \int \exd^{d+2}x \; \sqrt{-g} \; \Bigl( L_\ssZ + L_{\rm mix} + L_\Psi + V_b \Bigr) \,,
\eea
where $D_\ssM \Psi := \partial_\ssM \Psi - i e Z_\ssM \, \Psi$, and the second line again defines the various $L_i$.

For later purposes it is useful to write $\sqrt{2} \; \Psi = \psi \, e^{i \Omega}$ and adopt a unitary gauge for which the phase, $\Omega$, is set to zero, though this gauge will prove to be singular at the origin of the vortex solutions we examine later. In this gauge the term $L_\Psi$ in $S_\ssV$ can be written
\be
 L_\Psi = D_\ssM \Psi^* D^\ssM \Psi = \frac12 \Bigl( \partial_\ssM \psi \, \partial^\ssM \psi + e^2 \psi^2 Z_\ssM Z^\ssM \Bigr)
\ee
and the potential becomes
\be
 V_b = \frac{\lambda}4 \, \Bigl( \psi^2 - v^2 \Bigr)^2 \,.
\ee
It is also useful to group the terms in the brane and bulk lagrangians together according to how many metric factors and derivatives appear, with
\bea \nn
 &&\phantom{OO}L_{\rm kin} :=  \frac12 \, g^{\ssM\ssN} \partial_\ssM \psi \, \partial_\ssN \psi \,, \qquad
 L_{\rm gge} := L_\ssA + L_\ssZ + L_{\rm mix} \\
 &&L_{\rm pot} := \Lambda + V_b \qquad
 \hbox{and} \qquad
 L_{\rm gm} := \frac12 \,e^2 \psi^2 \, g^{\ssM \ssN} Z_\ssM Z_\ssN \,.
\eea
For this system the field equations for the two Maxwell fields are
\be \label{checkAeq}
 \frac{1}{\sqrt{-g}} \, \partial_\ssM \Bigl[ \sqrt{-g} \Bigl( A^{\ssM \ssN} + \varepsilon \, Z^{\ssM \ssN} \Bigr) \Bigr] = 0 \,,
\ee
and
\be \label{Z0eq}
 \frac{1}{\sqrt{-g}} \, \partial_\ssM \Bigl[ \sqrt{-g} \Bigl( Z^{\ssM \ssN} + \varepsilon \, A^{\ssM \ssN} \Bigr) \Bigr] = e^2 \Psi^2 Z^\ssN  \,.
\ee
The scalar field equation in unitary gauge becomes
\be \label{Psieom}
 \frac{1}{\sqrt{-g}} \, \partial_\ssM \Bigl( \sqrt{-g} \; g^{\ssM \ssN} \partial_\ssN \psi \Bigr) = e^2 \psi Z_\ssM Z^\ssM + \lambda \, \psi \Bigl(\psi^2 - v^2 \Bigr) \,,
\ee
while the Einstein equations can be written in their trace-reversed form
\be \label{TrRevEin}
 \cR_{\ssM \ssN} = - \kappa^2 X_{\ssM \ssN} \,,
\ee
where $X_{\ssM \ssN} := T_{\ssM \ssN} - (1/d) \, T \, g_{\ssM \ssN}$ and the stress-energy tensor is
\bea
 T_{\ssM\ssN} &=&  \partial_\ssM \psi \, \partial_\ssN \psi + e^2 \psi^2 Z_\ssM Z_\ssN + A_{\ssM\ssP} {A_\ssN}^\ssP + Z_{\ssM \ssP} {Z_\ssN}^\ssP \\
  && \qquad + \frac{\varepsilon}2 \, \Bigl( A_{\ssM\ssP} {Z_\ssN}^\ssP + Z_{\ssM \ssP} {A_\ssN}^\ssP \Bigr) - g_{\ssM \ssN} \Bigl( L_{\rm kin} + L_{\rm gm} + L_{\rm pot} + L_{\rm gge} \Bigr) \,.\nn
\eea

\subsection{Symmetry ans\"atze}
\label{subsec:ansatz}

We seek vortex solutions for which the brane/vortex sector describes energy localized along a time-like $d$-dimensional subspace, with nontrivial profiles in the two transverse dimensions. Accordingly, our interest is in configurations that are maximally symmetric in the $d$ dimensions (spanned by $x^\mu$) and axially symmetric in the 2 `transverse' dimensions (spanned by $y^m$).

We take the fields to depend only on the proper distance, $\rho$, from the points of axial symmetry, and assume the only nonzero components of the gauge field strengths lie in the transverse two directions: $A_{mn}$ and $Z_{mn}$. We choose the metric to be of warped-product form
\be \label{productmetric}
 \exd s^2 = g_{\ssM \ssN} \, \exd x^\ssM \exd x^\ssN = g_{mn} \, \exd y^m \exd y^n + g_{\mu\nu} \, \exd x^\mu \exd x^\nu \,,
\ee
with
\be \label{warpedprod}
 g_{mn} = g_{mn}(y) \qquad \hbox{and} \qquad
 g_{\mu\nu}  =  W^2(y) \, \check  g_{\mu\nu}(x) \,,
\ee
where $\check  g_{\mu\nu}$ is the maximally symmetric metric on $d$-dimensional de Sitter, Minkowski or anti-de Sitter space. The corresponding Ricci tensor is $\cR_{\ssM \ssN} \, \exd x^\ssM \exd x^\ssN = \cR_{\mu\nu} \, \exd x^\mu \exd x^\nu + \cR_{mn} \, \exd y^m \exd y^n$, and is related to the Ricci curvatures, $\check  R_{\mu\nu}$ and $R_{mn}$, of the metrics $\check  g_{\mu\nu}$ and $g_{mn}$ by
\be
 \cR_{\mu\nu} = \check  R_{\mu\nu} + g^{mn} \Bigl[ (d-1) \partial_m W \partial_n W + W \nabla_m \nabla_n W \Bigr] \, \check  g_{\mu\nu} \,,
\ee
and
\be \label{cR2vsR2}
 \cR_{mn} = R_{mn} + \frac{d}{W} \; \nabla_m  \nabla_n W \,,
\ee
where $\nabla$ is the 2D covariant derivative built from $g_{mn}$. We work with axially symmetric 2D metrics, for which we may make the coordinate choice
\be \label{xdmetric}
  g_{mn} \, \exd y^m \exd y^n = A^2(r) \, \exd r^2 + B^2(r) \, \exd \theta^2  =  \exd \rho^2 + B^2(\rho) \, \exd \theta^2  \,,
\ee
where the proper radial distance satisfies $\exd \rho = A(r) \exd r$. With these choices the field equation simplify to the following system of coupled nonlinear ordinary differential equations.

\medskip\noindent{\em Gauge fields}

\medskip\noindent
The gauge field equations become
\be \label{Acheckeom}
  \left( \frac{ W^d \check A_\theta'}{B} \right)' = 0 \,,
\ee
and
\be \label{Zeom}
 \frac{1 - \varepsilon^2}{BW^d} \, \left( \frac{ W^d Z_\theta'}{B} \right)' = \frac{e^2 \psi^2 Z_\theta}{B^2}  \,,
\ee
where primes denote differentiation with respect to proper distance, $\rho$, and we define the mixed gauge field,
\be \label{AZcheckA}
  \check A_{\ssM} :=  A_{\ssM} + \varepsilon \, Z_{\ssM} \,.
\ee

Notice that the off-diagonal contribution to $L_{\rm gge}$ vanishes when this is expressed in terms of $\check A_\ssM$ rather than $A_\ssM$, since
\be  \label{Lggemixed}
 L_{\rm gge} = L_\ssA + L_\ssZ + L_{\rm mix} = \check L_\ssA + \check L_\ssZ \,,
\ee
where
\be
 \check L_\ssA := \frac14 \, \check A_{mn} \check A^{mn}
 \quad \hbox{and} \quad
 \check L_\ssZ := \frac14 \, (1-\varepsilon^2)  Z_{mn}Z^{mn} \,.
\ee
Notice also that \pref{Zeom} has the same form as it would have had in the absence of the $A-Z$ mixing, \pref{AZcheckA}, provided we make the replacement $e^2 \to \hat e^2$, with
\be
 \hat e^2 := \frac{e^2}{1 - \varepsilon^2} \,.
\ee
Clearly stability requires the gauge mixing parameter must satisfy $\varepsilon^2 < 1$ and semi-classical methods require us to stay away from the upper limit.

\medskip\noindent{\em Scalar field}

\medskip\noindent
The field equation for $\psi(\rho)$ similarly simplifies to
\be \label{Psieom2}
 \frac{1}{BW^d} \, \Bigl( BW^d \, \psi' \Bigr)' = e^2 \psi \left( \frac{Z_\theta}{B} \right)^2 + \lambda \, \psi \Bigl(\psi^2 - v^2 \Bigr) \,.
\ee

\medskip\noindent{\em Einstein equations}

\medskip\noindent
The nontrivial components of the matter stress-energy become
\be \label{Tmunurhotot}
 T_{\mu\nu}  = - g_{\mu\nu} \; \varrho_{\rm tot}  \,, \qquad
 {T^\rho}_\rho = \cZ - \cX  \qquad \hbox{and} \qquad
 {T^\theta}_\theta = -( \cZ + \cX )  \,,
\ee
where
\be
 \varrho :=  L_{\rm kin} + L_{\rm gm} + L_{\rm pot} + L_{\rm gge} \,,
\ee
and
\be
 \cX := L_{\rm pot} - L_{\rm gge} \qquad \hbox{and} \qquad
 \cZ := L_{\rm kin} - L_{\rm gm} \,.
\ee
In later sections it is useful to split $\varrho = \varrho_{\rm loc} + \check \varrho_\ssB$, $\cX = \cX_{\rm loc} + \check \cX_\ssB$ and $\cZ = \cZ_{\rm loc}+ \cZ_\ssB$ into vortex and bulk parts, which we do as follows:
\bea \label{StressEnergyVBsplit}
 &&\check \varrho_\ssB := \Lambda + \check L_\ssA \,, \qquad
 \varrho_{\rm loc} := L_{\rm kin} + L_{\rm gm} + V_b + \check L_\ssZ \nn\\
 &&\check \cX_\ssB := \Lambda - \check L_\ssA \,, \qquad
 \cX_{\rm loc} := V_b - \check L_\ssZ \\
 \hbox{and} \quad
 &&\cZ_\ssB := 0 \,, \qquad\qquad\;\;
 \cZ_{\rm loc} := L_{\rm kin} - L_{\rm gm} = \cZ  \,. \nn
\eea

The components of the trace-reversed Einstein equations governing the $d$-dimensional on-vortex geometry therefore become
\be \label{4DTrRevEin}
 \cR_{\mu\nu} = - \kappa^2 X_{\mu\nu}
 = - \frac{2}{d} \; \kappa^2\cX \; g_{\mu\nu}  \,,
\ee
of which maximal symmetry implies the only nontrivial combination is the trace
\be \label{avR4-v1}
 \cR_{(d)} := g^{\mu\nu} \cR_{\mu\nu} = \frac{\check  R}{W^2} + \frac{d}{BW^d} \Bigl( BW' W^{d-1} \Bigr)'
 = -2\kappa^2 \cX \,,
\ee
and we use the explicit expression for $\cR_{(d)}$ in terms of $\check  R$ and $W$. The components dictating the 2-dimensional transverse geometry similarly are $\cR_{mn} = - \kappa^2 X_{mn}$, which has two nontrivial components. One can be taken to be its trace
\be \label{avR2}
 \cR_{(2)} := g^{mn} \cR_{mn} = R + d \left( \frac{W''}{W} + \frac{B'W'}{BW} \right) = - \kappa^2 {X^m}_m = - 2 \kappa^2 \left[  \varrho - \left( 1 - \frac{2}{d} \right) \cX \right] \,,
\ee
and the other can be the difference between its two diagonal elements
\be
 {\cG^\rho}_\rho - {\cG^\theta}_\theta = {\cR^\rho}_\rho - {\cR^\theta}_\theta = - \kappa^2 \left( {T^\rho}_\rho - {T^\theta}_\theta \right) \,.
\ee
Writing out the curvature and stress energy shows this last equation becomes
\be \label{newEinstein}
 \frac{B}{W} \left( \frac{W'}{B} \right)' = - \frac{2}{d} \; \kappa^2 \cZ \,.
\ee

\medskip\noindent{\em Other useful combinations of Einstein equations}

\medskip\noindent
Other linear combinations of the Einstein equations are not independent, but are sometimes more useful. The first is the $(\theta\theta)$ component of the trace-reversed Einstein equation  ${\cR^\theta}_\theta = - \kappa^2  {X^\theta}_\theta $ which reads
\be \label{XthetathetaeEinstein}
 \frac{ (B' W^d)' }{BW^d} = - \kappa^2  \left[ \varrho - \cZ - \left( 1 - \frac{2}{d} \right) \cX \right] = - 2\kappa^2 \left( L_{\rm gm} + L_{\rm gge} + \frac{\cX}{d} \right)  \,.
\ee
A second useful form is the $(\rho\rho)$ Einstein equation, ${\cG^\rho}_\rho = - \kappa^2 {T^\rho}_\rho$, which is special in that all second derivatives with respect to $\rho$ drop out. This leaves the following `constraint' on the initial conditions for the integration in the $\rho$ direction:
\bea \label{constraint}
   d \left( \frac{B'W'}{BW} \right) + \frac{\check  R}{2W^2} + \frac{d(d-1)}{2} \left( \frac{W'}{W} \right)^2 &=& \kappa^2 \Bigl( \cZ - \cX \Bigr) \nn\\
   &=& \kappa^2 \Bigl( L_{\rm kin} - L_{\rm gm} - L_{\rm pot} + L_{\rm gge} \Bigr) \,.
\eea

\subsection{Scales and hierarchies}
\label{subsec:scales}

Before solving these field equations, we first briefly digress to summarize the relevant scales that appear in their solutions. The fundamental parameters of the problem are the gravitational constant, $\kappa$; the gauge couplings, $e$ (for $Z_\ssM$) and $g_\ssA$ (for $A_\ssM$); the scalar self-coupling, $\lambda$, and the scalar vev $v$. These have the following engineering dimensions in powers of mass:
\be
 \left[ \kappa \right] = 1-D/2 \,, \qquad
 \left[ e \right] = \left[ g_\ssA \right] = 2-D/2 \,, \qquad
 \left[ \lambda \right] = 4-D \,, \quad \hbox{and} \quad
 \left[ v \right] = D/2-1 \,.
\ee
To these must be added the dimensionless parameter, $\varepsilon$, that measures the mixing strength for the two gauge fields.

In terms of these we shall find that the energy density of the vortex is of order $e^2 v^4$ and this is localized within a region of order
\be
 r_\varv = \frac{1}{ev} \,.
\ee
The effective energy-per-unit-area of the vortex is therefore of order $e^2 v^4 r_\varv^2  = v^2$. These energies give rise to $D$-dimensional curvatures within the vortex of order $1/L_\varv^2 = \kappa^2 e^2 v^4$ and integrated dimensional gravitational effects (like conical defect angles) of order $\kappa^2 v^2$. We work in a regime where $\kappa v \ll 1$ to ensure that the gravitational response to the energy density of the vortex is weak, and so, for example, defect angles are small and $L_\varv \gg r_\varv$.

By contrast, far from the vortex the curvature scale in the bulk turns out to be of order $1/r_\ssB^2$ where
\be
 r_\ssB = \frac{\cN \kappa}{g_\ssA} \,,
\ee
and $\cN$ is a dimensionless measure of the total amount of $A_\ssM$ flux that threads the compact transverse dimensions. Since our interest is in the regime where the vortex is much smaller than the transverse dimensions we throughout assume $r_\varv \ll r_\ssB$ and so
\be
  \frac{g_\ssA}{e\cN} \ll \kappa v \ll 1 \,.
\ee

\section{Isolated vortices}
\label{section:isolatedvortex}

We now describe some solutions to the above field equations, starting with the local properties of an isolated vortex within a much larger ambient bulk geometry. Our goal is to relate the properties of the vortex to the asymptotic behaviour of the bulk fields and their derivatives outside of (but near to) the vortex itself, with a view to using these as matching conditions when replacing the vortex with an effective codimension-2 localized object. These matching conditions are then used in later sections to see how a system of several vortices interact with one another within a compact transverse geometry. To this end we regard the field equations as to be integrated in the radial direction given a set of `initial conditions' at the vortex centre.

\subsection{Vortex solutions}
\label{subsec:vortexsoln}

For vortex solutions the vortex scalar vanishes at $\rho = 0$, and the vortex fields approach their vacuum values, $\psi \to v$ and\footnote{In unitary gauge.} $Z_\ssM \to 0$, at large $\rho$. Because we work in the regime $\kappa v \ll 1$ these solutions closely resemble familiar Nielsen-Olesen solutions \cite{NOSolns} in the absence of gravitational fields. Our analysis in this section reduces to that of \cite{OtherGVs} in the limit of no gauge mixing, $\varepsilon = 0$, and a trivial bulk.

The asymptotic approach to the far-field vacuum values can be understood by linearizing the field equations about their vacuum configurations, writing $\psi = v + \delta \psi$ and $Z_\theta = 0 + \delta Z_\theta$. We find in this way that both $\delta \psi$ and $\delta Z_\ssM$ describe massive particles, with respective masses given by
\be
 m^2_\ssZ = \hat e^2 v^2  \quad \hbox{and} \quad
 m^2_\Psi = 2\lambda v^2 \,.
\ee
From this we expect the approach to asymptopia to be exponentially fast over scales of order $r_\ssZ = m_\ssZ^{-1}$ and $r_\Psi = m_\Psi^{-1}$. Indeed this expectation is borne out by explicit numerical evaluation.

Notice the two vortex scales are identical, $r_\varv := r_\ssZ = r_\Psi$, in the special BPS case, defined by $\hat \beta = 1$ where
\be
 \hat \beta := {\hat e^2}/{2\lambda} \,,
\ee
and so the BPS case satisfies $\hat e^2 = 2 \lambda$. For convenience we also define $\beta = e^2 / 2 \lambda = (1 - \varepsilon^2) \hat \beta$.

\subsubsection*{Boundary conditions near the origin}

We start with a statement of the boundary conditions to be imposed at $\rho = 0$, which express that the transverse metric, $g_{mn}$, is locally flat and that all vectors (and so in particular the gradients of all scalars) must vanish there. For the metric functions we therefore impose the conditions
\be \label{WBBC}
 W(0) = W_0 \,, \quad W'(0) = 0  \qquad \hbox{and} \qquad
 B(0) = 0 \quad \hbox{and} \quad B'(0) = 1
\,.
\ee
We can choose $W_0 = 1$ by rescaling the $d$-dimensional coordinates, but this can only be done once so the {\em change}, $\Delta W$, between the inside and the outside of the vortex (or between the centres of different vorticies) is a physical thing to be determined by the field equations. Similarly, for the vortex scalar we demand
\be \label{psiBC}
 \psi(0) = \psi'(0) = 0 \,,
\ee
or we could also trade one of these for the demand that $\psi \to v$ far from the vortex core.

Nonsingularity of the bulk gauge field-strengths implies they must take the form
\be
 A_{mn} = f_\ssA \, \epsilon_{mn} \,, \quad
 Z_{mn} = f_\ssZ \, \epsilon_{mn} \quad \hbox{and so} \quad
 \check A_{mn} = \check f_\ssA \, \epsilon_{mn} \,,
\ee
where $\epsilon_{\rho\theta} = \sqrt{g_2} = B$ is the volume form for the 2D metric $g_{mn}$. Since $\check A_{mn}$ is nonsingular we know $\check f_\ssA$ is regular at $\rho = 0$ and so because $B(\rho) \simeq \rho$ near $\rho = 0$ we see that $\check A_{\rho\theta} \propto \rho$ near the origin. Consequently, in a gauge where $\check A_\ssM \, \exd x^\ssM = \check A_\theta(\rho) \, \exd \theta$ we should expect $\check A_\theta = \cO(\rho^2)$ near the origin.

Naively, the same should be true for the vortex gauge fields $A_\ssM$ and $Z_\ssM$, however the gauge transformation required to remove the phase everywhere from the order parameter $\Psi = \psi e^{i \Omega}$ ({\em i.e.} to reach unitary gauge) is singular at the origin, where $\Psi$ vanishes and so $\Omega$ becomes ambiguous. Consequently in this gauge $Z_\theta$ (and so also $A_\ssM$) does not vanish near the origin like $\rho^2$. Instead because in this gauge $Z_\ssM \to 0$ far from the vortex we see that flux quantization demands that
\be \label{ZBC}
   -\frac{2\pi n}{e} = \Phi_\ssZ(\rho < \rho_\varv) := \oint_{\rho = \rho_\varv} Z = 2\pi \int_0^{\rho_\varv} \exd \rho \, \partial_\rho Z_\theta = 2\pi \Bigl[ Z_\theta(\rho_\varv) - Z_\theta(0) \Bigr] = -2\pi Z_\theta(0)\,,
\ee
where $n$ is an integer, and we choose $\rho = \rho_\varv$ to be far enough from the vortex that $Z_\ssM \to 0$ there. We therefore ask $Z_\theta$ to satisfy the boundary condition:
\be
 Z_\theta(0)  = \frac{n}{e} \quad \hbox{and so therefore} \quad  A_\theta(0)  = -\frac{n \varepsilon}{e} \,,
\ee
where the second equality follows from $\check A_\theta (0) = 0$.

\subsubsection*{Vortex solutions}

\begin{figure}[t]
\centering
\includegraphics[width=\textwidth]{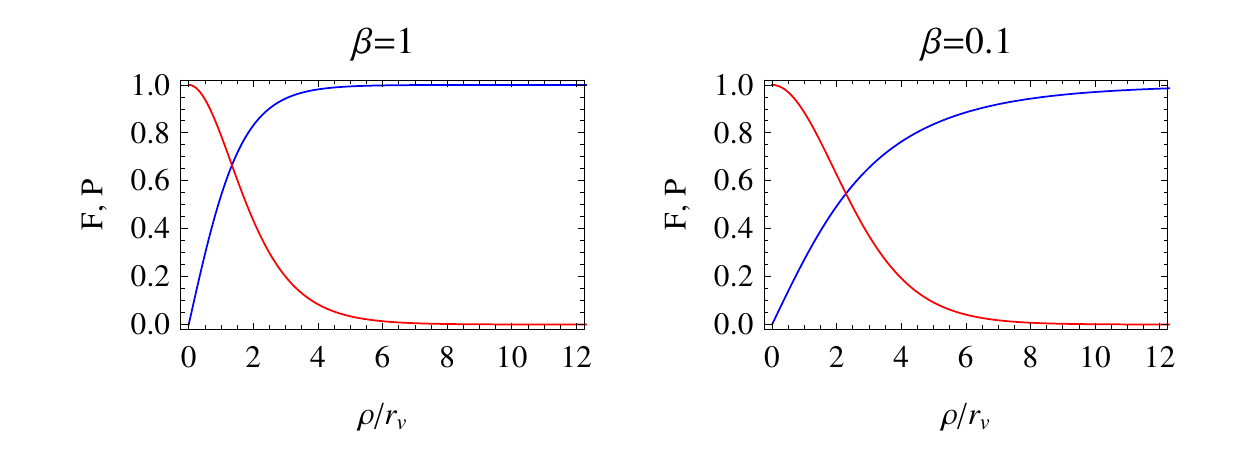}
\caption{A comparison of BPS and non-BPS vortex profiles on a flat background for differing values of $\hat \beta = \hat e^2/(2\lambda)$. The (blue) profile vanishing at the origin is the scalar profile $F$ and the (red) profile that decreases from the origin is the vector profile $P$. To find the profiles in flat space we set $B = \rho$ and $W=1.$  The left plot uses $\hat \beta = 1$ and the right plot uses $\hat \beta = 0.1$, with this being the only parameter that controls vortex profiles in flat space. }
\label{fig:flatprofiles}
\end{figure}

It is convenient to normalize the vortex fields
\be
 Z_\theta = \frac{n}{e} \; P(\rho) \qquad \hbox{and} \qquad
 \psi = v \; F(\rho) \,
\ee
so that $F = 1$ corresponds to the vacuum value $\psi = v$, while the boundary conditions at $\rho = 0$ become
\be
 F(0) = 0 \,, \quad P(0) = 1 \,;
\ee
the vacuum configuration in the far-field limit is
\be
 F(\infty) = 1 \,, \quad P(\infty) = 0 \,.
\ee
In terms of $P$ and $F$ the $Z_\ssM$ field equations boil down to
\be \label{Peq}
 \frac{1}{BW^d} \left( \frac{ W^d \, P'}{ B} \right)' = \frac{\hat e^2 v^2 F^2 P}{B^2} \,,
\ee
while the $\psi$ equation reduces to
\be \label{Feq}
 \frac{1}{BW^d} \Bigl( BW^d \; F' \Bigr)' = \frac{P^2 F}{B^2} + \lambda v^2\, F\left( F^2 - 1 \right) \,.
\ee
Although closed form solutions to these are not known, they are easily integrated numerically for given $B$ and $W$, and the results agree with standard flat-space results when $B = \rho$ and $W = 1$. See, for example, Fig.~\ref{fig:flatprofiles}.

\subsubsection*{BPS special case}

In the special case where $W = 1$ and $\hat e^2 = e^2/(1-\varepsilon^2) = 2 \lambda$ (and so $\hat \beta = 1$), eqs.~\pref{Peq} and \pref{Feq} are equivalent to the first-order equations,\footnote{The simplicity of these equations is understood in supersymmetric extensions of these models, since supersymmetry can require $e^2 = 2\lambda$ and the vortices in this case break only half of the theory's supersymmetries.}
\be \label{BPSeqs}
 B F' = n  F P  \qquad \hbox{and} \qquad
 \frac{n  P'}{\hat e B} = \sqrt{\frac{\lambda}{2}} \; v^2  \left(  F^2 - 1 \right) \,.
\ee
We show later that $W = 1$ also solves the Einstein equations when $\hat e^2 = 2 \lambda$ and so this choice provides a consistent solution to all the field equations in this case.

When eqs.~\pref{BPSeqs} and $W = 1$ hold, they also imply
\be
 L_{\rm kin} = \frac12 \, (\partial \psi)^2 =
  \frac{e^2}2 \, \psi^2 Z_\ssM Z^\ssM =  L_{\rm gm}\,,
\ee
and
\be \label{cLzeqVb}
 \check L_\ssZ := \frac14 \, (1-\varepsilon^2)  Z_{mn}Z^{mn} = \frac{\lambda}{4}  ( \psi^2 - v^2 )^2 = V_b \,,
\ee
which further imply that the vortex contributions to $\cZ$ and $\cX$ cancel out,
\be
 \cZ = L_{\rm kin} - L_{\rm gm} = 0 \quad
 \hbox{and} \quad
 \cX_{\rm loc} = V_b - \check L_\ssZ = 0 \,,
\ee
leaving only the bulk contribution to $\cX$:
\be
 \cX = \check \cX_\ssB = \Lambda - \check L_\ssA \,.
\ee
As can be seen from eq.~\pref{newEinstein}, it is the vanishing of $\cZ$ that allows $W = 1$ to solve the Einstein equations. Finally, the vortex part of the action evaluates in this case to the simple result
\be
 \cT_\varv := \frac{1}{\sqrt{- \check  g } } \int \exd^2y \, \sqrt{-g}\Bigl[ L_\Psi + V_b + \check L_\ssZ \Bigr]  = 2\pi \int \exd \rho \, B \; \Bigl[ L_\Psi + V_b + \check L_\ssZ \Bigr] = \pi n v^2 \,.
\ee

\begin{figure}[t]
\centering
\includegraphics[width=\textwidth]{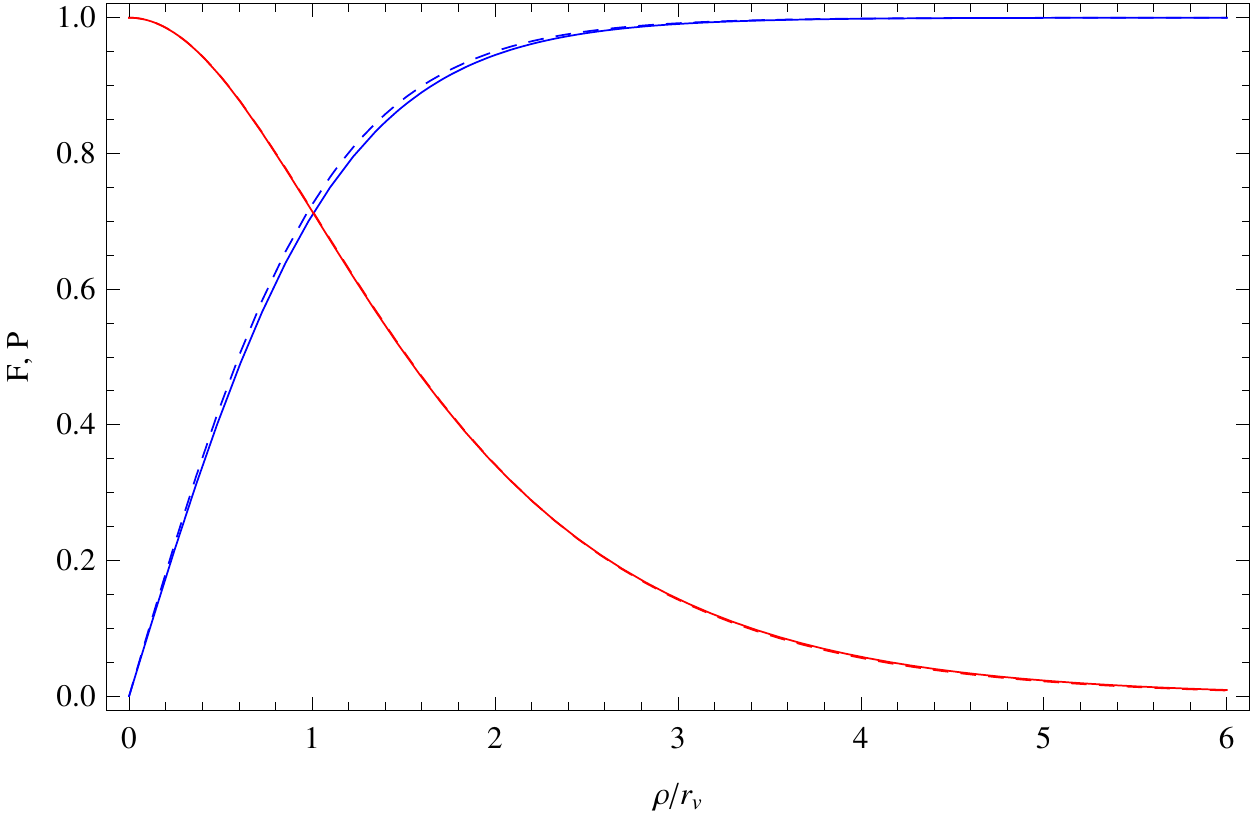}
\caption{A comparison of the profiles $F$ and $P$ for the vortex in flat space (dashed curves) and the full gravitating vortex solution (solid lines). For each case the (blue) profile that vanishes at the origin is the scalar profile $F$ and the (red) profile that decreases from the origin is the vector profile $P$. The parameters used in the plot are $d=4$, $\varepsilon = 0.3$, $\beta = 3$, $Q = 0.01 \, e v^2$, $\Lambda = Q^2/2$, $\kappa v = 0.6$ and $\check  R = 0$ with the same values of $\beta$ and $\varepsilon$ chosen for the non-gravitating solution.}
\label{fig:gravprofiles}
\end{figure}

\subsubsection*{Bulk equations}

To obtain a full solution for a vortex coupled to gravity we must also solve the bulk field equations for $W$, $B$ and $\check A_{\rho\theta}$. The simplest of these to solve is the Maxwell equation, \pref{Acheckeom}, whose solution is
\be \label{Acheckeomsoln}
  \check A_{\rho\theta} =  \frac{QB}{W^d}  \,,
\ee
where $Q$ is an integration constant. This enters into the Einstein equations, \pref{avR4-v1}, \pref{avR2} and \pref{newEinstein}, through the combination $\check L_\ssA = \frac12 (Q/W^d)^2$.

\begin{figure}[t]
\centering
\includegraphics[width=\textwidth]{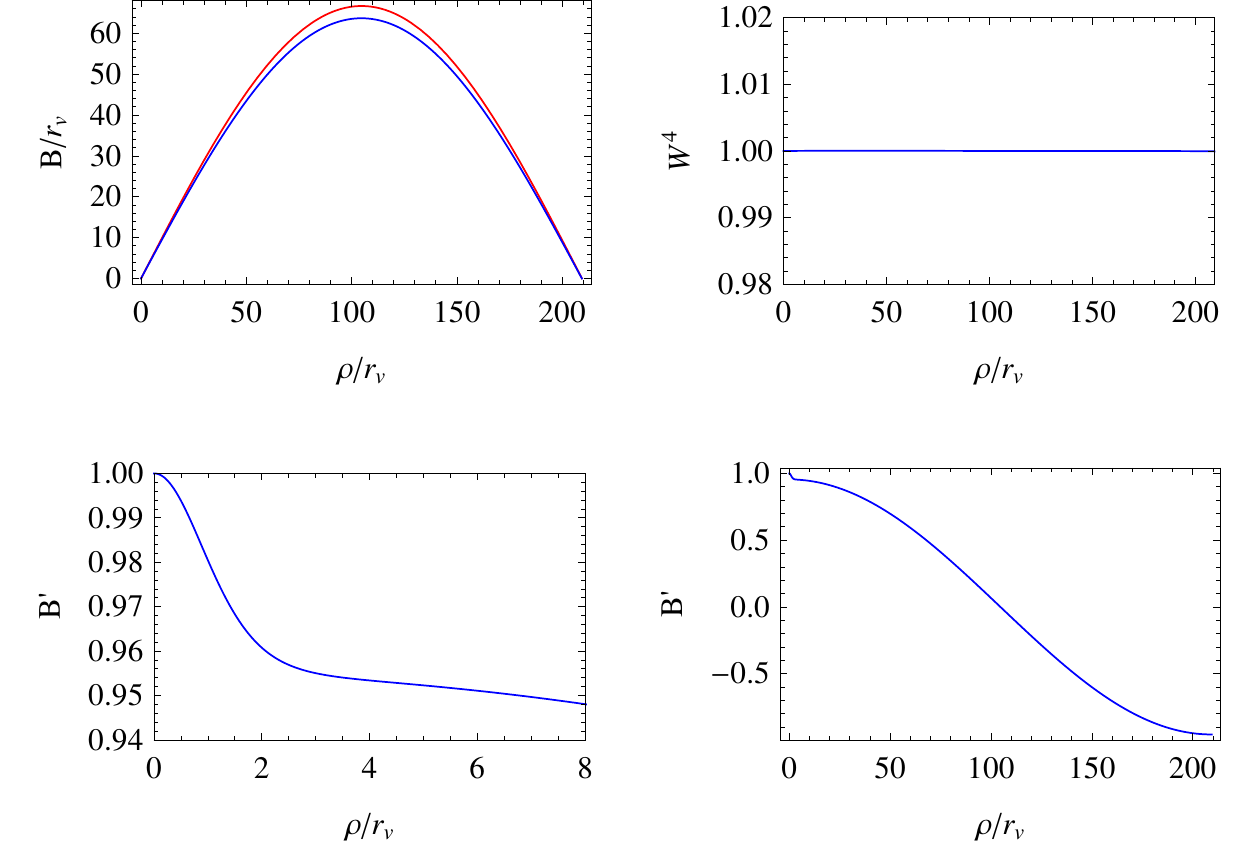}
\caption{These plots illustrate the bulk geometry for BPS vortices ($\beta = 1$) with parameters $d=4$,  $\varepsilon = 0$, $\beta = \hat \beta = 1$, $Q = 0.05 \, ev^2$, $\Lambda = Q^2/2$ and $\kappa v = 0.3$ (which also imply $\check  R = 0$). In the top left plot, the solution for $B$ is plotted (in blue) below the (red) metric function $B_{\rm sphere}$ of a sphere with radius $r_\ssB = (200/3) r_\varv  .$ The presence of a vortex does not change the size of the bulk (since the full solution for $B$ still vanishes at $\rho = \pi r_\ssB $) and the metric function $B$ is still approximately spherical with $B \approx 0.95 \times B_{\rm sphere}$ for these parameters. The top right plot shows that when $\beta = 1$ and $\Lambda = Q^2 / 2$, a constant warp factor solves the field equations. The bottom left plot shows that the derivative of the metric function $B' \approx 0.95$ outside of the vortex core, at $\rho \gsim 4 r_\varv$. The bottom right plot shows that $B' \approx -0.95$ at the pole which lies opposite to the vortex core, indicating the presence of a conical singularity at that pole.}
\label{fig:bulksoln}
\end{figure}

These can be numerically integrated out from $\rho = 0$, starting with the boundary conditions \pref{WBBC} (for which we choose $W_0 = 1$), \pref{psiBC} and \pref{ZBC}, provided that the curvature scalar, $\check  R$, for the metric $\check  g_{\mu\nu}$ is also specified.  Once this is done all field values and their derivatives are completely determined by the field equations for $\rho > 0$ and one such solution is shown in Fig.~\ref{fig:bulksoln}. As we shall see, many useful quantities far from the vortex depend only on certain integrals over the vortex profiles, rather than their detailed form.

\subsection{Integral relations}

Our main interest in later sections is in how the vortices affect the bulk within which they reside, and this is governed by the boundary conditions they imply for the metric --- {\em i.e.} on quantities like $W$, $W'$, $B$, $B'$ and $\check R$ --- as well as for other bulk fields exterior to, but nearby, the vortex. In particular, simple integral expressions exist for derivatives of bulk fields --- {\em e.g.} $W'$ and $B'$ --- in this near-vortex limit, and we pause here to quote explicit expressions for these.

\begin{figure}[t]
\centering
\includegraphics[width=0.55\textwidth]{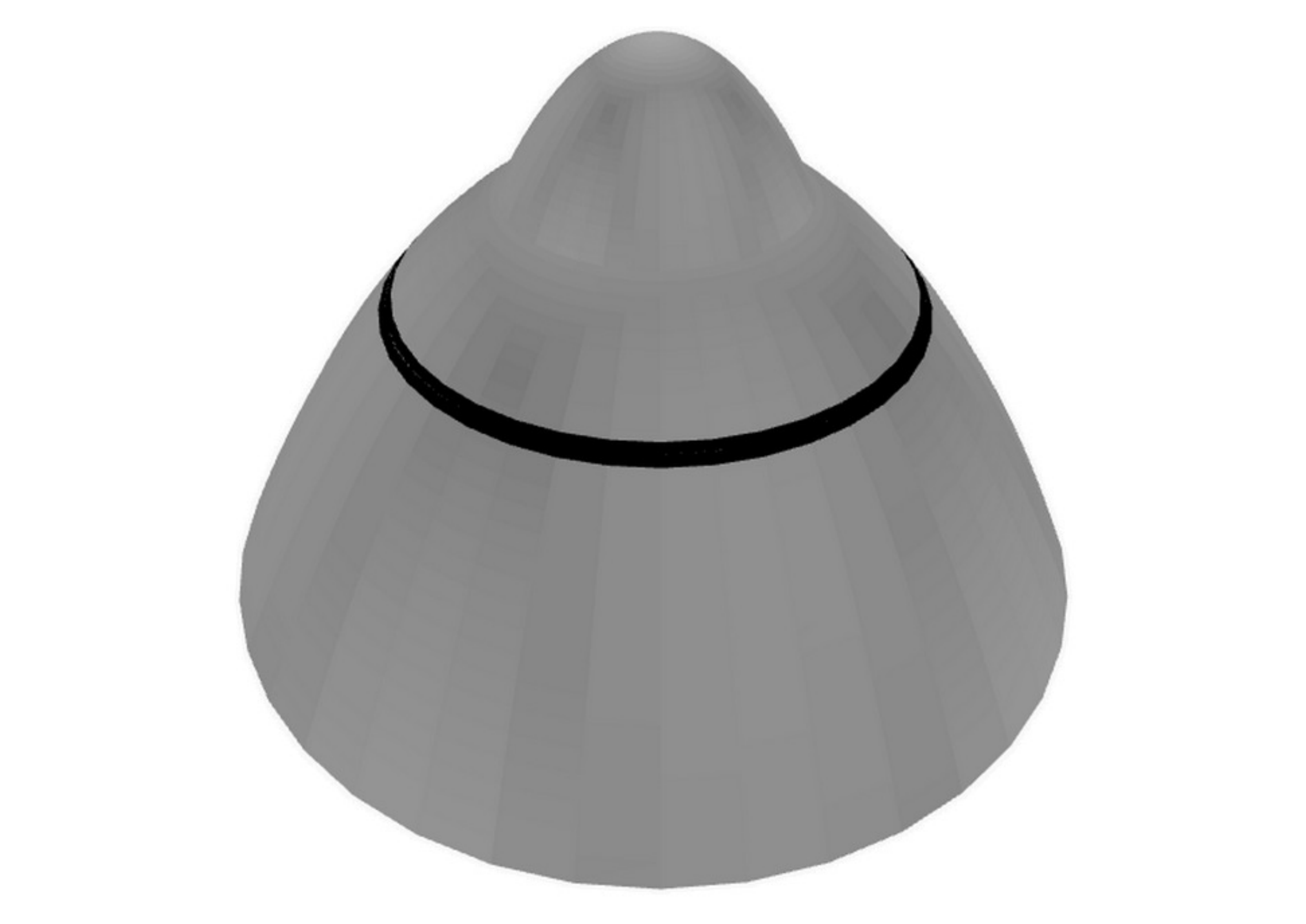}
\caption{An illustration of the matching done at $\rho = \rho_\varv.$ The light grey surface is a cartoon of the bulk geometry. The bump on top of the surface represents the localized modifications to the approximately spherical bulk geometry that arise due to the vortex. The dark ring represents the circle at $\rho = \rho_\varv$ that lies sufficiently far outside the vortex that its fields are exponentially suppressed, but close enough to the vortex so that that its proper distance from the pole is still $\cO(r_\varv).$}
\label{fig:cap}
\end{figure}

For instance, imagine integrating the Einstein equation, \pref{avR4-v1}, over the transverse dimensions out to a proper distance $\rho = \rho_\varv \simeq \cO(r_\varv)$ outside of (but not too far from) the vortex (see Figure \ref{fig:cap}). This gives
\be \label{intWval}
 d \,B  W^d \; \partial_\rho \ln W \Bigr|_{\rho=\rho_\varv} = \left[ B \Bigl( W^d \Bigr)' \right]_{\rho=\rho_\varv} = - \frac{1}{2\pi} \Bigl\langle 2\kappa^2 \cX + W^{-2} \check  R \Bigr\rangle_{\varv} \,,
\ee
where we introduce the notation
\be
 \left\langle \cO \right\rangle_\varv := \frac{1}{\sqrt{- \check  g } } \int\limits_{X_\varv} \exd^2x \, \sqrt{-g} \;  \cO = 2\pi \int_0^{\rho_\varv} \exd \rho \, B W^d \; \cO  \,,
\ee
and use the boundary condition $W'(0) = 0$ at the vortex centre. This identifies explicitly the specific combination of vortex quantities relevant for specifying $W'$ just outside the vortex.

A second integral relation of this type starts instead with the $(\theta\theta)$ component of the trace-reversed Einstein equation: ${\cR^\theta}_\theta = - \kappa^2 {X^\theta}_\theta$, or \pref{XthetathetaeEinstein}, which integrates to give
\be \label{BprmWdint}
 \Bigl( B' W^d \Bigr)_{\rho = \rho_\varv} = 1 - \frac{\kappa^2}{2\pi} \left\langle \varrho - \left( 1 - \frac{2}{d} \right) \cX - \cZ \right\rangle_{\varv} \,,
\ee
given the boundary condition $B'W^d \to 1$ as $\rho \to 0$. This can be used to infer the implications of the vortex profiles on $B'$ just outside the vortex.

For many purposes our interest is in the order of magnitude of the integrals on the right-hand sides of expressions like \pref{intWval} or \pref{BprmWdint} and these sometimes contain a surprise. In particular, naively one might think the integrals on the right-hand sides would generically be order $v^2$ and so would contribute at order $\kappa^2 v^2$ to the quantities on the left-hand sides. Although this is true for $\varrho$, the surprise is that the quantities $\langle \cX \rangle_\varv$ and $\langle \cZ \rangle_\varv$ can be much smaller than this, being suppressed by powers of $r_\varv/r_\ssB$ when the vortex is much smaller than the transverse space, $r_\varv \ll r_\ssB$, and this has important implications for how vortices influence their surroundings.

One way to understand this suppression is to evaluate explicitly the suppressed quantities in the flat-space limit, where it can be shown (for instance) that the vortex solutions described above imply $\langle \cX \rangle_{\varv\,{\rm flat}} = 0$. Appendix \ref{appsec:SEConservation} proves this as a general consequence of stress-energy conservation (or hydrostatic equilibrium) within the vortex, with the vortex dynamically adjusting to ensure it is true. (Alternatively, the vanishing of $\langle \cX \rangle_{\varv}$ on flat space can also be derived as a consequence of making the vortex action stationary with respect to rescalings of the size of the vortex.) More generally, for curved geometries we find numerically that in the generic situation when $r_\varv \sim r_\ssB$ all terms in \pref{intWval} are similar in size and not particularly small, but this is no longer true once a hierarchy in scales exists between the size of the vortex and that of the transverse dimensions. In particular, as shown in Appendix \ref{appsec:SEConservation}, for solutions where $\check  R$ is $1/r_\ssB^2$ suppressed the vortex dynamically adjusts also to suppress $\langle \cX \rangle_\varv$ by powers of $1/r_\ssB$.

The next sections provide several other ways to understand this suppression, associated with the constraints imposed by the Bianchi identities on the left-hand sides of near-vortex boundary conditions.

\subsection{Near-vortex asymptotics}
\label{subsec:nearvortex}

Because the vortex fields, $\delta \psi = \psi - v$ and $Z_\ssM$, fall off exponentially they can be neglected to exponential accuracy `outside' of the vortex; {\em i.e.} at distances $\rho_\varv  \gsim r_\varv \sim 1/ev$. The form for the metric functions $B$ and $W$ are then governed by the Einstein equations with only bulk-field stress-energy. This section describes the approximate form taken by these bulk solutions outside of the vortex sources, but not far outside (in units of the bulk curvature radius, say).

We do so in two steps. We first solve for the bulk fields external to an isolated vortex in an infinite transverse space. We then find approximate asymptotic solutions for vortices sitting within compact spaces, under the assumption that the compact space is much larger than the transverse vortex size and that the region of interest for the solutions is very close to the vortex: $r_\varv \lsim \rho \ll r_\ssB$.

\subsubsection*{Infinite transverse space}

We start with solutions where the space transverse to the vortex is not compact, since these should share many features of the bulk sufficiently close to a vortex residing within a large but finite transverse space. Concretely, the merit of seeking non-compact solutions is that the boundary conditions at infinity are fixed and determine many of the bulk integration constants. As seen in \S\ref{section:interactions}, compact spaces are more complicated from this point of view because these constants are instead set dynamically by the adjustment of the various vortices to each other's presence. But those near-vortex boundary conditions that are dictated by vortex properties should not care about distant details of the vortex environment, and so can be explored most simply within an isolated-vortex framework.

To find isolated solutions we first write the Einstein equations in the exterior region $\rho > \rho_\varv$ where the vortex fields can be neglected:
\be \label{einsteinthetatheta}
 \frac{( W^d B' )'}{W^d B} = -\kappa^2 \left[ \left( \frac{d-1}{d} \right) Q^2 W^{-2d} + \frac{2 \Lambda}{d} \right] \,,
\ee
and
\be \label{einsteindiff}
 d B \left( \frac{W'}{B} \right)' = 0 \,,
\ee
and
\be \label{einsteinrhorho}
 W^{-2} \check  R + \frac{( (W^d)' B)'}{W^d B} = \kappa^2 \left( Q^2 W^{-2d } - 2 \Lambda \right) \,.
\ee
In this section (and only this section) we assume the transverse space does not close off, so $B(\rho) > 0$ strictly for all values $\rho > \rho_\varv$.

Integrating \pref{einsteindiff} from $\rho_\varv$ to arbitrary $\rho > \rho_\varv$ gives
\be \label{kcoeff}
 \frac{W'}{B}  = \frac{W'_\varv}{B_\varv} = k \,,
\ee
where $k$ is an integration constant and a $\varv$ subscript indicates that the bulk field is evaluated at $\rho = \rho_\varv$. Evaluating this at infinity tells us $k = 0$ if we demand $W'$ vanishes there. More generally, if $k \ne 0$ and $B$ monotonically increases then $|W|$ must diverge at infinity, even if $B$ (and so $W'$) is bounded. Since $B > 0$ this excludes $k < 0$ since this would imply $W$ vanishes at finite $\rho > \rho_\varv$. If we also exclude $W \to \infty$ as $\rho \to \infty$ then we must have $k = 0$, for which integrating eq.~\pref{kcoeff} implies $W = W_\varv$ is constant everywhere outside the vortex.

Using this result in eq.~\pref{einsteinthetatheta} gives
\be
  \frac{B''}{B} = -\kappa^2 \left[ \left( \frac{d-1}{d} \right) Q^2 W_\varv^{-2d} + \frac{2 \Lambda}{d} \right] =: Y_d \,,
\ee
where the constancy of the right hand side (which we call $Y_d$) implies the transverse dimensions have constant curvature. Solving this for $B$ in the region $\rho > \rho_\varv$ gives elementary solutions whose properties depend on the sign of $Y_d$ :
\begin{itemize}
 \item $Y_d = - 1/\ell^2 < 0$: This implies $B$ is a linear combination of $\sin(\rho/\ell)$ and $\cos(\rho/\ell)$ and so eventually passes through zero to pinch off at some $r_\star > \rho_\varv$. This gives a compact transverse space, which we discard in this section.
 \item $Y_d = + 1/\ell^2 > 0:$ This implies $B$ is a linear combination of $\sinh(\rho/\ell)$ and $\cosh(\rho/\ell)$ and so increases exponentially for large $\rho$. This corresponds to a vortex sitting within an infinite-volume transverse hyperbolic space with curvature radius $\ell$.
 \item $Y_d = 0:$ This forces $B'' = 0$ which gives the flat solution $B = B_\varv + B_\varv'(\rho - \rho_\varv)$.
\end{itemize}

A flat transverse space is found by tuning the bulk cosmological constant such that $Y_d = 0$, and so
\be \label{Lambdachoice}
   \Lambda  = - \frac{1}{2} \left( d-1 \right) Q^2 W_\varv^{-2d} < 0 \,.
\ee
Having $\Lambda$ more negative than this gives a hyperbolic transverse space and more positive gives a compact transverse space. Evaluating \pref{einsteinrhorho} at the position $\rho = \rho_\varv$, using \pref{Lambdachoice} and constant $W = W_\varv$ then gives
\be \label{Rsolved}
  W_\varv^{-2} \check  R  = \kappa^2 \left( Q^2 W_\varv^d - 2 \Lambda \right) = d \kappa^2 Q^2 W^{-2d}_\varv = - 2 \kappa^2 \Lambda \left( \frac{d}{d-1} \right) > 0 \,,
\ee
which in our curvature conventions represents a strictly anti-de Sitter (AdS) geometry for the directions parallel to the vortex whenever the transverse directions are noncompact.

As argued in more detail in \S\ref{subsec:Exactsolns}, in general the 2D curvature scale, $R = \pm 2/\ell^2$, and the $d$ D scale, $\check R$, are independent functions of the two dimensionful parameters: $1/r_\Lambda^2 \propto \kappa^2\Lambda$ and $1/r_\ssA^2 \propto \kappa^2 Q^2$. Of special interest is the one-parameter subspace of configurations for which either $R$ or $\check R$ vanish, and the above shows that the case $\check R = 0$ necessarily involves finite-volume transverse dimensions while flat transverse space ($R = 0$) implies an AdS on-vortex geometry, so the two subspaces intersect only as both $r_\Lambda$ and $r_\ssA$ tend to infinity ({\em ie} for $\Lambda,\, Q \to 0$).

It is the constancy of $W = W_\varv$ in the bulk for isolated vortices that reflects something general about vortices: that $W' \to 0$ in the near-vortex limit. Indeed, although \S\ref{section:Rugbyballgeom} gives explicit compact solutions with $W' \ne 0$ in the bulk, in all cases $W'$ approaches zero in the immediate vicinity of the small source vortices. This carries implications for the integrated vortex stress-energy, such as $\langle \cX \rangle_\varv$. Using $W'_\ssV = 0$ in \pref{intWval} allows us to write
\be
 2 \kappa^2 \la \cX \ra_\varv = - \check  R \, \la W^{-2} \ra_\varv  \,,
\ee
which is useful because it shows that $\langle \cX \rangle_\varv$ is very generally suppressed by two powers of a curvature scale, being order $\rho_\varv^2/\check\ell^2 \ll 1$ if $\check R \sim 1/\check\ell^2 \ll 1/\rho_\varv^2$. We expect this result also to hold for vortices situated within compact transverse dimensions.

\subsubsection*{Asymptotic forms}

We next return to the case of real interest: small vortices situated within a much larger (but compact) transverse space. In general, the presence of a vortex introduces apparent singularities into the bulk geometry whose properties are dictated by those of the vortex. These singularities are only apparent because they are smoothed out once the interior structure of the vortex is included, since the geometry then responds to the stress-energy of the vortex interior. This section characterizes these singularities more precisely with a view to relating them to the properties of the source vortices.

\begin{figure}[h]
\centering
\includegraphics[width=0.55\textwidth]{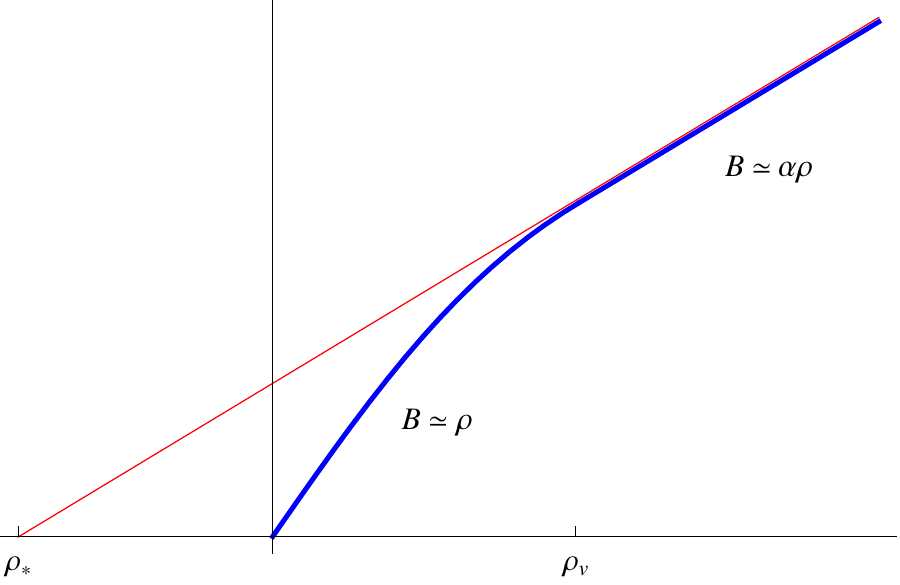}
\caption{A cartoon illustration of the definition of $\rho_\star$. The (blue) metric function $B$ increases linearly away from the origin with unit slope $B(\rho) \approx \rho$. Outside of the vortex $\rho \gsim \rho_\varv$ the solution is also linear in $\rho$ but with $B(\rho) \approx \alpha \rho$. The straight (red) line extrapolates this exterior behaviour to the point, $\rho = \rho_\star$,  where the external $B$ would have vanished if the vortex had not intervened first. }
\label{fig:rhostar}
\end{figure}

One way to characterize the position of the apparent singularity is to define it to occur at the point where the expression for $B_{\rm bulk}(\rho)$ obtained using only the bulk field equations would vanish: $B_{\rm bulk}(\rho_\star) = 0$ (see Figure \ref{fig:rhostar}). Here $\rho_\star$ is of order the vortex size, and need not occur precisely at $\rho = 0$ (despite the boundary condition $B(0) = 0$ inside the vortex) because $B_{\rm bulk}$ is found by solving only the bulk field equations without the vortex fields.

The nature of the singularity at $\rho = \rho_\star$ is most simply described by expanding the bulk field equations in powers of proper distance, $\hat\rho = \rho - \rho_\star$, away from the apparent singularity,
\bea \label{powerformsapp}
 W &=& W_0 \left( \frac{\hat\rho}{r_\ssB} \right)^w + W_1 \left( \frac{\hat\rho}{r_\ssB} \right)^{w+1} + W_2 \left( \frac{\hat\rho}{r_\ssB} \right)^{w+2} + \cdots  \,, \nn\\
 B &=& B_0 \left( \frac{\hat\rho}{r_\ssB} \right)^b  + B_1 \left( \frac{\hat\rho}{r_\ssB} \right)^{b+1} + B_2 \left( \frac{\hat\rho}{r_\ssB} \right)^{b+2} + \cdots  \,.
\eea
where $r_\ssB$ is again a scale of order the bulk curvature scale. It is the leading powers, $b$ and $w$, that describe potential singularity, and their form is constrained by the bulk field equations. In particular, as shown in Appendix~\ref{KasnerApp}, the leading terms in the expansion of the Einstein equations around $\hat\rho = 0$ imply that $w$ and $b$ satisfy the two Kasner conditions\footnote{Our treatment here follows closely that of \cite{6DdS}, which in turn is based on the classic BKL treatment of near-singularity time-dependence \cite{BKL}. } \cite{Kasner}:
\be \label{branekasner}
 dw + b = 1 \qquad \text{and} \qquad d w^2 + b^2 = 1 \,.
\ee
The last of these in turn implies $w$ and $b$ must reside within the intervals
\be
 | w | \le \frac{1}{\sqrt{d}} \qquad \text{and} \qquad |b| \le 1 \,.
\ee

The Kasner solutions have precisely two solutions: either $w = 0$ and $b = 1$ (as is true for flat-space solutions) or $w = 2/(1+d)$ and $b = (1-d)/(1+d)$. Since we know that a non-gravitating vortex lives in a geometry with $w = 0$ and $b = 1$, this is also the root we must use in the weak-gravity limit $( \kappa v )^2 \ll 1$. This describes a conical singularity if $B'(\rho = \rho_\star) \ne 1$.

\begin{figure}[t]
\centering
\includegraphics[width=\textwidth]{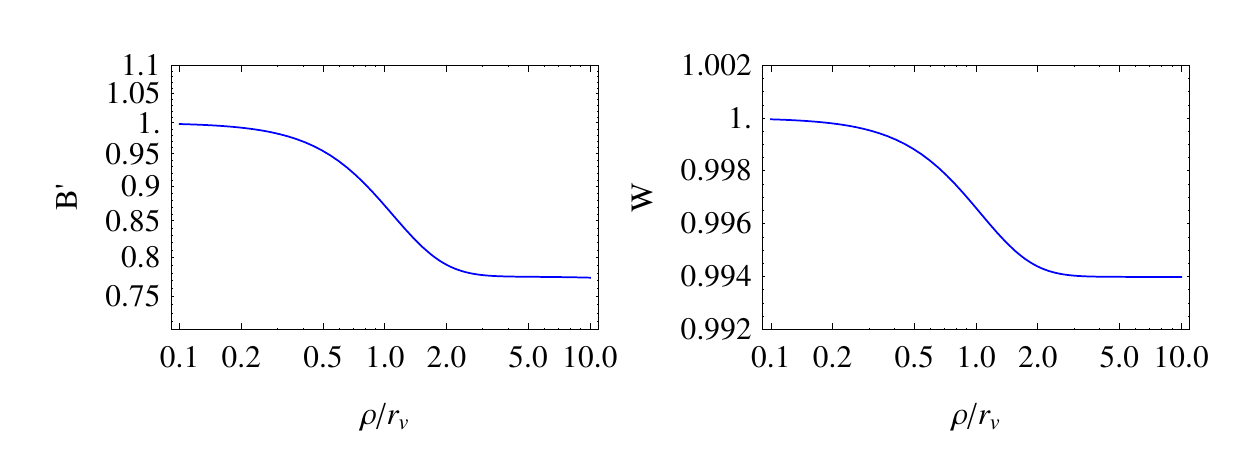}
\caption{Log-log plots of the the near vortex geometry for parameters $d=4$, $\beta = 3$, $\varepsilon = 0.3$, $Q = 0.01 \, ev^2$, $\Lambda = Q^2/2$, $\kappa v = 0.6$ and $\check  R = 0.$ The bulk in this case has a radius of $r_\ssB = (500/3) r_\varv$. Outside of the vortex $\rho \gsim r_\varv$ the geometry exhibits Kasner-like behaviour $B' \approx \alpha \ne 1$ and $W' \approx 0$. }
\label{fig:logprofiles}
\end{figure}

The field equations also dictate all but two of the remaining coefficients, $B_i$ and $W_i$, of the series solution. For instance eq.~\pref{newEinstein} applied outside the vortex implies $W' = k B$ for constant $k$. This implies $W_1 = 0$ and $W_2 = \frac{1}{2} \, k \,  \alpha\,r_\ssB^2$ and so on, giving
\bea \label{powerformsolnW'kB}
 W &=& W_\star + \left( \frac{k \alpha}{2} \right) \hat\rho^2 + \cdots  \,, \nn\\
 B &=& \alpha \, \hat \rho  + \cdots \,,
\eea
where $W_\star = \lim_{\rho \to \rho_\star} W$. For any such a singular point we therefore have the boundary conditions
\be
 \lim_{\rho \to \rho_\star} W^\prime = 0  \qquad \text{and} \qquad  \lim_{\rho \to \rho_\star} B^\prime =: \alpha = {\rm const} \,,
\ee
as is indeed found in detailed numerical integrations (see Figure \ref{fig:logprofiles}).

It is the slope $B_\star^\prime = \alpha$ and $W_\star$ (where we affix $W(0) = 1$ within the vortex and so are not free to again choose $W_\star = 1$ elsewhere) that convey the properties of the vortex to the bulk, and so should be governed by vortex properties, such as by boundary conditions like \pref{intWval} or \pref{BprmWdint}, rather than by bulk field equations. Notice that we expect both $W_\star - 1$ and $\alpha -1$ to be of order $\kappa^2 v^2$ (see below) and so if $W_2 = \frac{1}{2} k \,  \alpha \, r_\ssB^2$ is $\cO(1)$ then we expect $k \simeq \cO(1/r_\ssB^2)$. This, in turn, implies that $W' \simeq \cO(r_\varv / r_\ssB^2)$ at any near-vortex point of order $r_\varv$ away from $\rho_\star$. For $r_\varv \ll r_\ssB$ we expect $W$ to be approximately constant in the near-vortex region exterior to the vortex, up to $\cO(\rho^2/r_\ssB^2)$ corrections. We also expect $B'$ to be similarly constant, up to $\cO(\rho/r_\ssB)$ corrections. These expectations are verified by explicit numerical integrations of the vortex/bulk profiles, such as in Fig.~\ref{fig:logprofiles}.

The explicit relation between $\alpha$ and vortex properties is set by near-vortex boundary conditions, such as \pref{intWval} or \pref{BprmWdint}. Using the series expansion to evaluate $W$ and $B$ at $\rho = \rho_\varv$,
\be
 W = W_\varv + W'_\ssV  (\rho - \rho_\varv) + \cdots  \qquad \hbox{and} \qquad
 B = B_\varv + \alpha \, ( \rho - \rho_\varv)  + \cdots \,,
\ee
where $W_\varv = W_\star + \frac{k \alpha}{2} (\rho_\varv - \rho_\star)^2 + \cdots$, while $W_\varv' = k B_\ssV = k \alpha (\rho_\varv - \rho_\star) + \cdots$ and so on. Inserting these into the left-hand side of eqs.~\pref{intWval} then gives
\be \label{intWvalmatch}
 d \,B_\varv  W_\varv^{d-1} W_\varv' = dk W_\star^{d-1} \alpha^2 \hat\rho_\varv^2 + \cdots = - \frac{1}{2\pi} \Bigl\langle 2\kappa^2 \cX + W^{-2} \check  R \Bigr\rangle_{\varv} \,,
\ee
which confirms that the vortex adjusts to make the right-hand side $\cO(r_\varv^2/r_\ssB^2)$. Similarly \pref{BprmWdint} becomes
\be \label{BprmWdintmatch}
 B_\varv' W_\varv^d = \alpha W_\star^d + \cdots = 1 - \frac{\kappa^2}{2\pi} \left\langle \varrho - \left( 1 - \frac{2}{d} \right) \cX - \cZ \right\rangle_{\varv} \,,
\ee
and so on.

\subsection{Effective description of a small vortex}
\label{subsec:vortexEFT}

If the vortex is much smaller than the transverse space then most of the details of its structure should not be important when computing how it interacts with its environment. Its dynamics should be well described by an effective $d$-dimensional action that captures its transverse structure in a multipole expansion.

The lowest-derivative `brane' action of this type that depends on the nontrivial bulk fields outside the vortex is $S_b =  \int \exd^d x \; \cL_b$ with
\be \label{dSeff}
 \cL_b = -\sqrt{-\gamma} \left[ T - \frac{\zeta}{d\,!} \, \epsilon^{\mu\nu\lambda\rho}  \tilde A_{\mu\nu\lambda\rho} + \cdots \right]_{\rho = \rho_b} = -\sqrt{-\gamma} \left[ T + \frac{\zeta}{2} \, \epsilon^{mn} A_{mn} + \cdots \right]_{\rho = \rho_b} \,,
\ee
where $\gamma$ denotes the determinant of the induced metric on the $d$-dimensional world-volume of the vortex centre of mass (which in the coordinates used here is simply $\gamma_{\mu\nu} = g_{\mu\nu}$ evaluated at the brane position). The tensor $\tilde A_{\mu\nu\lambda\rho} := \frac12 \, \epsilon_{\mu\nu\lambda\rho mn} A^{mn}$ is proportional to the $D$-dimensional Hodge dual of the bulk field strength; a quantity that can be invariantly integrated over the $d$-dimensional world-volume of the codimension-2 vortex. All unwritten terms covered by the ellipses in \pref{dSeff} involve two or more derivatives.

The dimensionful effective parameters $T$ and $\zeta$ respectively represent the vortex's tension and localized flux, in a way we now make precise. To fix them in terms of the properties of the underlying vortex we perform a matching calculation; computing their effects on the bulk fields and comparing this to the parallel calculation using the full vortex solution. To do this we must be able to combine the $d$-dimensional action \pref{dSeff} with the $D$-dimensional action, $S_\ssB$, for the bulk fields.

To make this connection we promote \pref{dSeff} to a $D$-dimensional action by multiplying it by a `localization' function, $\delta(y)$, writing the $D$-dimensional lagrangian density as
\be \label{Leffdelta}
 \cL_{\rm tot} = \cL_\ssB(g_{\ssM\ssN},A_\ssM) + \cL_b (g_{\ssM\ssN},A_\ssM) \, \delta(y) \,.
\ee
Here $\cL_\ssB$ is as given in \pref{SB} and $\delta(y)$ is a delta-function-like regularization function that has support only in a narrow region around the vortex position $\rho = \rho_b$, normalized so that $\int_\ssV \exd^2 y \; \delta(y) = 1$. Although we can regard $\delta(y)$ as being independent of the $d$-dimensional metric, $g_{\mu\nu}$, and gauge field, $A_\ssM$, we {\em cannot} consider it to be independent of the transverse metric, $g_{mn}$, because $\delta(y)$ must depend on the proper distance from the vortex.

Much of the trick when matching with regularized delta-functions is to avoid questions that involve making assumptions about the detailed $g_{mn}$-dependence of the brane action. This is most awkward when calculating the brane's gravitational response, but we show below how to infer this response in a model-independent way that does not make ad-hoc assumptions about how $\delta(y)$ is regulated.

\subsubsection*{Gauge-field matching}

We start with the determination of the coupling $\zeta$ from the vortex's gauge-field response.

To determine $\zeta$ we compute the contribution of $S_b$ to the gauge field equation, which becomes modified to
\be
 \partial_m \Bigl( \sqrt{-g} \; A^{mn} \Bigr) + \frac{\delta S_b}{\delta A_n} = \partial_m \Bigl[ \sqrt{-g} \left( A^{mn} + \zeta \, \epsilon^{mn} \, \frac{\delta(y)}{\sqrt{g_2}} \right) \Bigr] = 0 \,.
\ee
This has solution
\be \label{Aeffeom}
 A_{\rho\theta} = \frac{QB}{W^d} - \zeta \, \epsilon_{\rho\theta} \, \frac{\delta(y)}{\sqrt{g_2}} = \frac{QB}{W^d} - \zeta  \, \delta(y) \,,
\ee
where $Q$ is an integration constant, and so --- when integrated over a transverse volume, $X_\varv$, completely containing the vortex --- gives the flux
\be
 \Phi_\ssA(X_\varv) = \int\limits_{X_\varv} \exd A = Q \int\limits_{X_\varv} \exd^2 y \left( \frac{B}{W^d} \right) - \zeta \,.
\ee
Comparing this to the vortex result in the full UV theory
\be
 \Phi_\ssA(X_\varv) = \check \Phi_\ssA(V) - \varepsilon \, \Phi_\ssZ(V)
 = Q \int\limits_{X_\varv} \exd^2 y \left( \frac{B}{W^d} \right) + \frac{2\pi n \varepsilon}{e} \,,
\ee
shows that $\zeta$ is given at the classical level by
\be \label{zetamatch}
 \zeta = -\frac{2\pi n \varepsilon}{e} \,.
\ee
Notice that this argument does not make use of any detailed properties of $\delta(y)$ beyond its normalization and independence of $A_m$.

\subsubsection*{Gauge-field back-reaction}

Before repeating this argument to match the tension, $T$, and determine the gravitational response, we first pause to draw attention to an important subtlety. The subtlety arises because the presence of localized flux causes the gauge field to back-react in a way that contributes to the localized energy density, in a manner similar to the way the classical Coulomb field back-reacts to renormalize the mass of a point charged particle.

To set up this discussion, notice that the effective lagrangian, \pref{dSeff}, can be regarded as the macroscopic contribution of the vortex part of the lagrangian regarded as a function of applied fields $A_m$ and $g_{\mu\nu}$. Consequently we expect the transverse average of \pref{Leffdelta} to give the same answer as the transverse average of the full lagrangian of the UV theory. Comparing the $A_m$-dependent and -independent terms of this average then suggests the identifications
\bea \label{Tzeta-UV}
 T \, W_b^d  &=& \Bigl\langle L_{\rm kin} + V_b + L_{\rm gm} + L_\ssZ \Bigr\rangle_\varv \nn\\
 \hbox{and} \quad \frac{\zeta}{2} \, W_b^d \, \epsilon^{mn} A_{mn} &=& \Bigl\langle L_{\rm mix} \Bigr\rangle_\varv = \frac{\varepsilon}{2} \Bigl \langle Z^{mn} A_{mn} \Bigr \rangle_\varv \,,
\eea
where $W_b = W(\rho_b)$ is the warp factor evaluated at the brane position, and the factors $W_b^d$ come from the ratio of $\sqrt{-\gamma}/\sqrt{- \check  g}$.

Now comes the main point. The existence of the localized piece in the solution, \pref{Aeffeom}, for $A_m$ has two related consequences in such a transverse average.
\begin{itemize}
\item First, evaluating the localized-flux term at the solution to the $A_{mn}$ field equation, \pref{Aeffeom}, shows that the localized component of $A_m$ renormalizes the tension,
   \be \label{LbevalA}
      W_b^d \left( T + \frac{\zeta}{2} \, \epsilon^{mn}A_{mn} \right)_{\rho=\rho_b} = W_b^d \left[ T + \frac{\zeta \, Q}{W_b^d} - \zeta^2 \left( \frac{\delta(y)}{B} \right)_{\rho=\rho_b} \right]  \,,
   \ee
   where this follows from taking $\delta(y)$ to be sufficiently peaked so that its integral can be treated like that of a Dirac delta-function.
   Notice that the last term in the last equality is singular as the vortex size goes to zero, requiring a regularization in order to be unambiguous. Such divergences are common for back-reacting objects with codimension-2 or higher, and are ultimately dealt with by renormalizing the action \pref{dSeff} even at the classical level \cite{ClassRenorm}.

   The $\zeta$-dependent part of this is to be compared with
   \be
     \Bigl\langle L_{\rm mix} \Bigr\rangle_\varv = -\frac{2\pi \varepsilon Q \, n}{e} - 2\varepsilon^2 \Bigl\langle L_\ssZ \Bigr\rangle_\varv\,,
   \ee
   which uses \pref{AZcheckA} and \pref{Acheckeomsoln} to evaluate the integration over $L_{\rm mix}$, and shows that the result agrees with \pref{LbevalA}, both on the value of the term linear in $Q$ (once the matching value, \pref{zetamatch}, for $\zeta$ is used) and by providing an explicit regularization of the singular $\cO(\varepsilon^2)$ term.
\item The second way the localized term in \pref{Aeffeom} contributes is by introducing a localized contribution to the Maxwell action, $L_\ssA$, which was naively not part of the vortex
    \bea
     \Bigl \langle L_\ssA \Bigr \rangle_\varv &=& \frac{Q^2}{2} \int\limits_{X_\varv} \exd^2 y \left( \frac{B}{W^d} \right) - W_b^d \left[ \frac { \zeta \, Q }{W_b^d} - \frac{\zeta^2}{2} \,  \left( \frac{\delta(y)}{B} \right)_{\rho=\rho_b} \right] \nn\\
     &=& \Bigl \langle \check L_\ssA \Bigr \rangle_\varv - W_b^d \left[ \frac { \zeta \, Q }{W_b^d} - \frac{\zeta^2}{2} \,  \left( \frac{\delta(y)}{B} \right)_{\rho=\rho_b} \right]  \,.
    \eea
    This exactly cancels the linear dependence on $Q$ in \pref{LbevalA}, and partially cancels the localized renormalization of the tension.
\end{itemize}

We see from this that the localized part of the gauge response to the brane action contributes a localized contribution to the bulk action (and energy density) that combines with the direct brane action in precisely the same way as happens microscopically from the mixing from $A_m$ to $\check A_m$ (see, for example, \pref{Lggemixed}). This suggests another useful notion of brane lagrangian, defined as the total localized contribution when $Q$ is fixed (rather than $A_m$), leading to
\be \label{endofhighlight}
 \check L_b := \check T \, W_b^d := \Bigl \langle L_{\rm kin} + V_b + L_{\rm gm} + \check L_\ssZ \Bigr \rangle_\varv
 = W_b^d \left[ T - \frac{\zeta^2}{2} \left( \frac{\delta(y)}{B} \right)_{\rho=\rho_b} \right] \,.
\ee
We see that the tension renormalizations described above --- associated with the $[\delta(y)/B]_{\rho_b}$ terms --- are the macroscopic analogs of the renormalization $e^2 \to \hat e^2 = e^2/(1 - \varepsilon^2)$ that occurs with the transition from $L_\ssZ$ to $\check L_\ssZ$ in the microscopic vortex picture.

Whether $L_b$ or $\check L_b$ is of interest depends on the physical question being asked. $L_b$ arises in deriving the brane contribution to the $A_m$ field equations, as above. But because it is $\check L_b$ that contains all of the brane-localized contributions to the energy, it plays a more important role in the brane's gravitational response (as we now explore in more detail).

\subsubsection*{On-brane stress energy}

With the above definitions of $L_b$ and $\check L_b$ in hand we now turn to the determination of the brane's local gravitational response. To determine the tension, $T$ (or $\check T$), we compute the $(\mu\nu)$ component of the Einstein equations (which we can do unambiguously because we know $\delta(y)$ does not depend on $g_{\mu\nu}$). We can do so using either $L_b$ or $\check L_b$ to define the brane action.

Using $L_b$ leads to the following stress energy
\be \label{branestressenergy}
 T^{\mu\nu}_{(b)} = \frac{2}{\sqrt{-g}} \, \left( \frac{\delta S_b}{\delta g_{\mu\nu}} \right) = - W_b^d \left( T + \frac{\zeta}{2} \, \epsilon^{mn}A_{mn} \right) \frac{\delta(y)}{\sqrt{g_2}} \; g^{\mu\nu} \,,
\ee
and so $\varrho$ becomes $\varrho = \Lambda + L_\ssA + \varrho_b$ with
\be \label{rhobdelta}
 \varrho_b = W^d_b \left( T + \frac{\zeta}{2} \, \epsilon^{mn}A_{mn} \right) \frac{\delta(y)}{\sqrt{g_2}} \,.
\ee
Alternatively, using $\check L_b$ leads to the stress energy
\be
 \check T^{\mu\nu}_{(b)} = \frac{2}{\sqrt{-g}} \, \left( \frac{\delta \check S_b}{\delta g_{\mu\nu}} \right) = - \check T \, W_b^d \; \frac{\delta(y)}{\sqrt{g_2}} \; g^{\mu\nu} \,,
\ee
and so $\varrho$ becomes $\varrho = \Lambda + \check L_\ssA + \check \varrho_b$ with
\be \label{rhobdeltacheck}
 \check \varrho_b = \check T \, W_b^d \; \frac{\delta(y)}{\sqrt{g_2}} =
  W_b^d \left[ T - \frac{\zeta^2}{2} \left( \frac{\delta(y)}{B} \right)_{\rho=\rho_b} \right] \frac{\delta(y)}{\sqrt{g_2}} \,.
\ee
In either case the {\em total} energy density is the same,
\be \label{rhobevalA}
 \left\langle \varrho \right\rangle_\varv = \Bigl \langle \Lambda + L_\ssA \Bigr \rangle_\ssV + W_b^d \left( T + \frac{\zeta}{2} \, \epsilon^{mn}A_{mn} \right)_{\rho=\rho_b} = \Bigl \langle \Lambda + \check L_\ssA \Bigr \rangle_\varv + W_b^d \, \check T  \,,
\ee
which is the analog of the microscopic statement \pref{StressEnergyVBsplit}
\be
  \bigl\langle \varrho  \bigr\rangle_\varv = \Bigl\langle \Lambda + L_\ssA  +  L_{\rm kin} + L_{\rm gm} + V_b + L_\ssZ + L_{\rm mix} \Bigr\rangle_\varv = \Bigl\langle \Lambda + \check L_\ssA  +  L_{\rm kin} + L_{\rm gm} + V_b + \check L_\ssZ \Bigr\rangle_\varv  \,.
\ee
The advantage of using \pref{rhobdeltacheck} rather than \pref{rhobdelta} is that $\check\varrho_b$ contains {\em all} of the brane-localized stress energy, unlike $\varrho_b$ which misses the localized energy hidden in $L_\ssA$.

\subsubsection*{IR metric boundary conditions}

A second important step in understanding the effective theory is to learn how the effective action modifies the field equations. So we restate here the general way of relating brane properties to near-brane derivatives of bulk fields \cite{BraneToBC}. The idea is to integrate the bulk field equations (including the brane sources) over a small region not much larger than (but totally including) the brane. For instance for a bulk scalar field, $\Phi$, coupled to a brane one might have the field equation
\be
 \Box \Phi + J_\ssB + j_b \, \delta(y) = 0 \,,
\ee
where $J_\ssB$ is the contribution of bulk fields that remains smooth near the brane position and $j_b$ is the localized brane source. Integrating this over a tiny volume surrounding the brane and taking its size to zero --- {\em i.e.} $\rho_\varv/r_\ssB \to 0$ --- then gives
\be
 \lim_{\rho_\varv \to 0} \Bigl \langle \Box \Phi \Bigr \rangle_\varv = 2\pi \lim_{ \rho_\varv \to 0} B_\varv W_\varv^d \, \Phi_\varv' = - \lim_{\rho_\varv \to 0} \Bigl \langle J_\ssB + j_b \, \delta (y) \Bigr\rangle_\varv = - j_b(\rho = \rho_b) \,,
\ee
where the assumed smoothness of $J_\ssB$ at the brane position ensures $\langle J_\ssB \rangle_\varv \to 0$ in the limit $\rho_\varv \to 0$. The equality of the second and last terms of this expression gives the desired relation between the near-brane derivative of $\Phi$ and the properties $j_b$ of the brane action.

Applying this logic to the Einstein equations, integrating over a tiny volume, $X_\varv$, completely enclosing a vortex gives
\be
   0 = \left \langle \frac{g^{\ssM \ssP} }{\sqrt{-g}} \frac{\delta S}{ \delta g^{\ssN \ssP} } \right \rangle_\varv = \left \langle \frac{g^{\ssM \ssP} }{\sqrt{-g}} \frac{\delta S_{\ssE \ssH} }{ \delta g^{\ssN \ssP} } \right \rangle_\varv + \left \langle \frac{g^{\ssM \ssP} }{\sqrt{-g}} \frac{\delta S_\ssM}{ \delta g^{\ssN \ssP} } \right \rangle_\varv
\ee
where we have split the action into an Einstein-Hilbert part $S_{\ssE \ssH}$ and a matter part $S_\ssM.$ This matter part can be further divided into a piece that is smooth at the brane position
\be
 \check S_\ssB = - \int \d^\ssD x \sqrt{-g} \left( \check L_\ssA + \Lambda \right) \,,
\ee
and one that contains {\em all} of the localized sources of stress energy,
\be
 \check S_b = - \int \d^\ssD x \sqrt{-g} \left( \frac{ \delta(y)}{\sqrt{g_2}} \right) \; \check T = - \int \d^d x \sqrt{- \gamma} \; \check T \,.
\ee

As above, for a sufficiently small volume, $X_\varv$, we need keep only the highest-derivative part of the Einstein-Hilbert term\footnote{Being careful to include the Gibbons-Hawking-York action \cite{GHY} on the boundary.}, since the remainder vanishes on integration in the limit $\rho_\varv \to 0.$ The $S_\ssM^\ssB$ term also vanishes in this limit, by construction, so the result becomes
\be \label{intermedmatching}
 0 = \frac{1}{2\kappa^2} \int \exd \theta \left[ \sqrt{-g} \Bigl( K^i{}_j - K \, \delta^i{}_j \Bigr) \right]^{\rho_\varv}_0 + \sqrt{-\check  g} \left\langle \frac{g^{i k} }{\sqrt{-g} } \frac{\delta \check S_b }{\delta g^{j k} }\right\rangle_\varv \, \quad \text{as} \quad \rho_\varv \to 0 \,,
\ee
where $i$ and $j$ run over all coordinates except the radial direction, $\rho$, and $K^{ij}$ is the extrinsic curvature tensor for the surfaces of constant $\rho$. To proceed, we assume that the derivative of the brane action is also localized such that its integral can be replaced with a quantity evaluated at the brane position
\be
 \left\langle \frac{g^{\ssN \ssP} }{\sqrt{-g} } \frac{\delta \check S_b }{\delta g_{\ssM \ssP} }\right\rangle_\varv =  \int\limits_{X_\varv} \d^2 y \,  \left( \frac{g^{\ssN \ssP}}{\sqrt{- \check  g} } \frac{\delta \check S_b}{\delta g^{\ssM \ssP}} \right) = \left( \frac{ g^{\ssN \ssP } }{\sqrt{- \check  g} }  \frac{ \delta \check S_b }{\delta g_b^{\ssM \ssP} } \right)_{\rho = \rho_b} \,.
\ee
The $b$ subscript in the functional derivative of the last term denotes that it is taken at the fixed point where $\delta(y)$ is localized, and so it contains no dependence on the bulk coordinates, and in particular no factors of $\delta(y)$. For example its $\mu\nu$ components read
\be \label{branemunuderiv}
 \frac{ \delta \check S_b }{\delta g^{\mu \nu}_b } = - \left. \frac{1}{2} \sqrt{-\gamma} \; \check T \, g_{\mu \nu} \right|_{\rho = \rho_b} \,.
\ee
However, at this point we remain agnostic about how to calculate the off-brane component $\delta \check S_b / \delta g_{\theta \theta}$. Returning to the matching condition \pref{intermedmatching} we have the final result
\be \label{diffmatching}
  \lim_{\rho_\varv \to 0} \int \exd \theta \left[ \sqrt{-g}  \Bigl( K^i{}_j - K \, \delta^i{}_j \Bigr) \right]^{\rho_\varv}_0  = - 2 \kappa^2   \left( g^{ik }\frac{ \delta \check S_b }{\delta g_b^{jk} }  \right)_{\rho = \rho_b} \,,
\ee
which can be explicitly evaluated for the geometries of interest.

\subsubsection*{Brane stress-energies}

We now turn to the determination of the off-brane components of the brane stress-energy. We can learn these directly by computing the left hand side of \pref{diffmatching} in the UV theory, before taking the limit $\rho_\varv \to 0$. We will first do this very explicitly for the $(\mu \nu)$ components of the brane stress-energy, and then proceed to deduce the off-brane components of the brane stress-energy.

\medskip\noindent{\em The $(\mu \nu)$ stress-energy}

\medskip\noindent
For the metric ansatz $\exd s^2 = W^2(\rho) \, \check  g_{\mu\nu} \, \exd x^\mu \exd x^\nu + \exd \rho^2 + B^2(\rho) \, \exd \theta^2$, the extrinsic curvature evaluates to $K_{ij} = \frac12 \, g_{ij}'$. This gives
\be
 K^{\mu\nu} = \frac{W'}{W} \, g^{\mu\nu}  \qquad \hbox{and} \qquad
 K^{\theta\theta} = \frac{B'}{B} \, g^{\theta\theta}  \,.
\ee
The trace of the $(\mu\nu)$ components of the condition \pref{diffmatching} therefore evaluates to
\be \label{metricmatchingmunu}
  \lim_{\rho_\varv \to 0} \left\{ W_\varv^d  B_\varv  \left[ (1-d) \left( \frac{W'_\varv}{W_\varv} \right) - \frac{B'_\varv}{B_\varv} \right] + 1 \right\} = - \frac{\kappa^2/\pi d}{ \sqrt{-\check  g} } \left( g^{\mu\nu} \, \frac{\delta \check S_b}{\delta g_b^{\mu\nu}} \right)_{\rho= \rho_b} = \frac{\kappa^2 \, W_b^d \, \check T}{2\pi} \,,
\ee
for which the limit on the left-hand side can be evaluated using the limit $B_\varv \to 0$ as $\rho_\varv \to 0$. The result shows that it is the renormalized tension, $\check T$, that determines the defect angle just outside the vortex,
\be \label{defectmatching}
  1 -  \alpha = \frac{\kappa^2 \,W_b^d \, \check T}{2 \pi}  \,.
\ee
This is the macroscopic analog of  \pref{BprmWdint}.

\begin{figure}[t]
\centering
\includegraphics[width=1\textwidth]{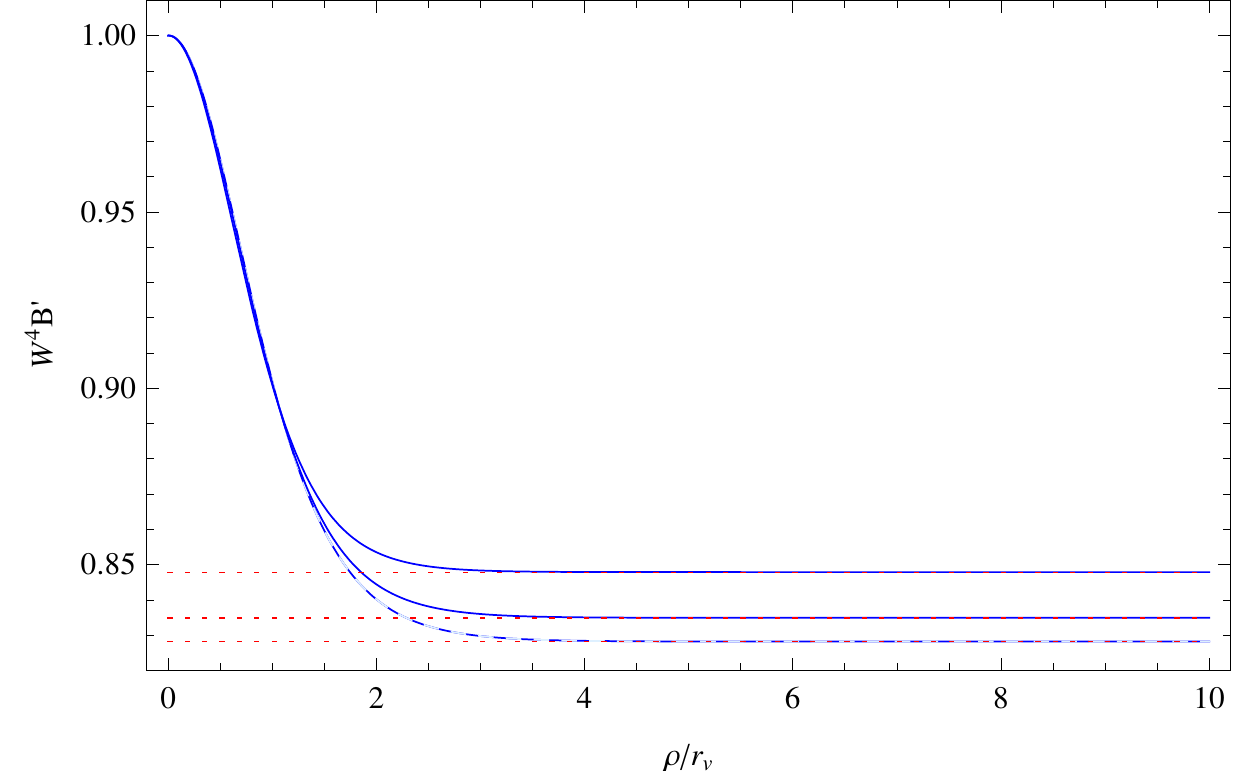}
\caption{A plot of defect angle matching in the region exterior but near to the vortex core. The solid (blue) lines represent the metric function $W^4 B'$ and the dotted (red) lines represent $1 - \kappa^2 \check T / 2 \pi$ computed independently for different values of $\varepsilon = \{-0.2, 0.2, 0.4, 0.6\}$ with the other parameters fixed at $d=4$, $\beta = 3$, $Q = 1.25 \times 10^{-4} \, ev^2$, $\Lambda = Q^2/2$, $\kappa v = 0.5$ and $\check  R = 0$. This size of the defect angle $B^\prime_\ssV \approx \alpha$ matches very well with $1 - \kappa^2 \check T / 2 \pi$ at $\rho = \rho_\varv \approx 4 r_\varv .$ The solutions for $W^4 B'$ overlap perfectly when $\varepsilon = \pm 0.2$, as indicated by the dashes in the line. This illustrates that the defect angle is controlled by $\check T$, and the linear dependence of the the defect angle on $\varepsilon$ is cancelled.
}
\label{fig:defect}
\end{figure}

\medskip\noindent{\em The $(\theta\theta)$ stress-energy}

\medskip\noindent
The $(\theta\theta)$ component of the metric matching condition, \pref{diffmatching}, evaluates to
\be \label{metricmatchingthth}
   \lim_{\rho_\varv \to 0} W^d_\varv B_\varv \left( \frac{W'_\varv}{W_\varv} \right) = \frac{\kappa^2/\pi d}{ \sqrt{-\check  g} }   \left( g^{\theta \theta} \, \frac{\delta \check S_b}{\delta g_b^{\theta \theta}} \right)_{\rho= \rho_b}  \,.
\ee
but at first sight this is less useful because the unknown $g_{mn}$ dependence of $\delta(y)$ precludes evaluating its right-hand side.
This problem can be side-stepped by using the constraint, eq.~\pref{constraint}, evaluated at $\rho = \rho_\varv$ (just outside the brane or vortex) to evaluate $W_\varv'/W_\varv = \cO(\rho_\varv/r_\ssB^2)$ (and so also the left-hand side of \pref{metricmatchingthth}) in terms of the quantities $B_\varv'/B_\varv = 1/\rho_\varv + \cdots$, $\check  R/W_\varv^2$ and $\cX_\ssB$. Once this is done we instead use the $(\theta\theta)$ matching condition to infer the $(\theta\theta)$ component of the vortex stress energy.

Solving the constraint, \pref{constraint}, for $W'/W$ at $\rho_\varv$ (just outside the vortex, where $\cZ = 0$ and $\cX = \cX_\ssB = \check \cX_\ssB$) gives
\bea
(d-1) \left( \frac{W'_\varv}{W_\varv} \right) &=& -  \frac{B'_\varv}{B_\varv} + \sqrt{ \left( \frac{B'_\varv}{B_\varv} \right)^2 - \left( 1 - \frac{1}{d} \right) \left( 2 \kappa^2 \cX_\ssB(\rho_\varv) + \frac{\check  R}{W_\varv^2} \right)} \nn\\
 &\simeq&  - \frac{1}{2} \left( 1 - \frac{1}{d} \right) \rho_\varv \left( 2 \kappa^2 \cX_\ssB(\rho_\varv) + \frac{\check  R}{W_\varv^2} \right) + \cdots \,,
\eea
where the root is chosen such that $W'_\varv/W_\varv$ vanishes if both $\check  R$ and $\cX_\ssB(\rho_\varv)$ vanish. With this expression we see that $B_\varv W_\varv^d (W_\varv'/W_\varv) \to 0$ as $\rho_\varv \to 0$, and so \pref{metricmatchingthth} then shows that
\be \label{ththzero}
\left( g^{\theta \theta} \, \frac{\delta \check S_b}{\delta g_b^{\theta \theta}} \right)_{\rho= \rho_b} = 0 \,,
\ee
for any value of $T$ (or $\check T$) and $\zeta$.

Notice that eq.~\pref{ththzero} is precisely what is needed to ensure $W_b' \to 0$ at the brane, as required by the Kasner equations \pref{branekasner} that govern the near-vortex limit of the bulk. Also notice that \pref{ththzero} would be counter-intuitive if instead one were to evaluate directly $\delta S_b/\delta g_{\theta\theta}$ by assuming $\delta(y)$ was metric independent and using the explicit metrics that appear within $\epsilon^{mn}A_{mn}$. What is missed by this type of naive calculation is the existence of the localized energy coming from the Maxwell action, $L_\ssA$, and its cancellation of the terms linear in $\zeta$ when converting $S_b$ to $\check S_b$.

\medskip\noindent{\em The $(\rho\rho)$ stress-energy}

\medskip\noindent
Although the $(\rho \rho)$ component of the extrinsic curvature tensor is not strictly well-defined, we can still consider the $(\rho \rho)$ components of the boundary condition in the following form
\be
 0 =  \lim_{\rho_\varv \to 0} \left \langle \frac{g^{\rho \rho} }{\sqrt{-g}} \frac{\delta S_{\ssE \ssH} }{ \delta g^{\rho \rho} } \right \rangle_\varv + \left( \frac{ g^{\rho \rho} }{\sqrt{- \check  g} }  \frac{ \delta \check S_b }{\delta g_b^{\rho \rho} } \right)_{\rho = \rho_b} \,.
\ee
By definition, we have
\be
\frac{g^{\rho \rho} }{\sqrt{-g}} \frac{\delta S_{\ssE \ssH} }{ \delta g^{\rho \rho} } = -\frac{1}{2 \kappa^2}\cG^\rho{}_\rho \,.
\ee
As noted in \pref{constraint}, this component of the Einstein tensor is contains only first derivatives of the metric field. It follows that
\be
\lim_{\rho_\varv \to 0} \left \langle \frac{g^{\rho \rho} }{\sqrt{-g}} \frac{\delta S_{\ssE \ssH} }{ \delta g^{\rho \rho} } \right \rangle_\varv =  -\frac{1}{2 \kappa^2} \left \langle \cG^\rho{}_\rho \right \rangle_\varv = 0
\ee
since metric functions and their first derivatives are assumed to be smooth. In this simple way, we once again use the Hamiltonian constraint to conclude that the off-brane component of the brane stress energy is vanishing
\be
 \left( \frac{ g^{\rho \rho} }{\sqrt{- \check  g} }  \frac{ \delta \check S_b }{\delta g_b^{\rho \rho} } \right)_{\rho = \rho_b} = 0 \,.
\ee
So both off-brane components of brane stress-energy vanish in the limit $\rho_\varv \to 0$, and from this we also infer that their sums and differences also vanish:
\be \label{XZbmatching}
 \cX_b = \cZ_b = 0 \,.
\ee
These results are the analog for the effective theory of the KK-suppression of $\langle \cX \rangle_\varv$ in the UV theory once $r_\varv \ll r_\ssB$. As a consequence in the effective theory
\be
 \cZ = 0 \qquad \hbox{and} \qquad  \cX = \check \cX_\ssB = \Lambda - \check L_\ssA \,.
\ee

\section{Compactification and interacting vortices}
\label{section:interactions}

We next turn to how several small vortices interact with one another and with their environment. In particular, if the flux in the transverse dimensions does not fall off quickly enough its gravitational field eventually dominates and drives $B(\rho)$ to zero for positive $\rho$, thereby pinching off and compactifying the two transverse dimensions. We explore in detail the situation of two small vortices situated at opposite sides of such a compact space.

For this part of the discussion it is more convenient to use the effective description of the vortices as codimension-2 branes than to delve into their detailed vortex substructure, though we do this as well to see how the effective description captures the full theory's low-energy behaviour. As we saw above, in the effective limit the vortex properties are encoded in the near-brane derivatives of the bulk fields (through the defect angle and localized flux). So to discuss brane interactions it is useful to start with the general solution to the bulk field equations outside the vortices, since it is the trading of the integration constants of this solution for the near-brane boundary conditions that expresses how brane properties back-react onto their environs.

\subsection{Integral relations}
\label{subsec:integrals}

Before delving into explicit solutions to the bulk field equations, it is worth first recording some exact results that can be obtained by applying the integrated forms of the field equations to the entire transverse space, and not just to a small region encompassing each vortex. In the UV theory these integrals simplify because all fields are everywhere smooth and so the integral over a total derivative vanishes. The same need not be true for the effective theory with point brane sources, since in principle fields can diverge at the brane locations. However we can then ask how the UV integral relations arise in the effective theory.

For instance if eq.~\pref{intWval} is integrated over the entire compact transverse space then its left-hand side integrates to zero, leaving the following exact relation for $\check  R$:
\be \label{intWvaltot}
 0 = \Bigl\langle 2\kappa^2 \cX \Bigr \rangle_{\rm tot} + \Bigl \langle  W^{-2} \Bigr\rangle_{\rm tot} \check  R = 2\kappa^2 \left( \langle \cX \rangle_{\rm tot} + \frac{ \check  R}{2\kappa^2_d} \right) \,.
\ee
Here the last equality uses the relation between $\kappa^2$ and its $d$-dimensional counterpart. This shows that it is $\langle \cX \rangle_{\rm tot}$ that ultimately determines the value of the on-brane curvature.

Eq.~\pref{intWvaltot} is particularly powerful in the effective theory, for which we have seen that the branes satisfy $\check \cX_b = 0$ and so $\cX = \check \cX_\ssB = \Lambda - \check L_\ssA$. In this case \pref{intWvaltot} shows us that it is really only through
\be \label{checkLAvac}
 \langle \check L_\ssA \rangle_{\rm tot} = 2\pi Q^2 \int_{\rm tot} \exd \rho \,\left( \frac{B}{W^d} \right)
\ee
that the brane properties determine the on-brane curvature, as they modify the functional form of $B$ and $W^d$ through boundary conditions, and $Q$ through flux quantization.

A second exact integral relation comes from integrating the $(\theta\theta)$ component of the trace-reversed Einstein equation, eq.~\pref{XthetathetaeEinstein}, over the transverse dimensions. Again the left-hand side integrates to zero leaving the constraint
\be
 \left\langle \varrho - \cZ - \left( 1 - \frac{2}{d} \right) \cX \right \rangle_{\rm tot} = 0 \,.
\ee
Combining this with \pref{intWvaltot} then implies
\be \label{tRvsvarrho}
 \check  R = - \left( \frac{2d}{d-2} \right) \, \kappa_d^2 \Bigl \langle \varrho - \cZ  \Bigr \rangle_{\rm tot} \,.
\ee

Notice that in $d$ dimensions Einstein's equations with a cosmological constant, $V_{\rm eff}$, have the form
\be
 \check  R_{\mu\nu} - \frac12 \, \check  R \, \check  g_{\mu\nu} = \kappa_d^2 \, V_{\rm eff} \, \check  g_{\mu\nu}  \,,
\ee
and so the scalar curvature satisfies
\be \label{tRvsVeff}
 \check  R = - \left( \frac{2d}{d-2} \right) \, \kappa_d^2 V_{\rm eff} \,.
\ee
Comparing this with eq.~\pref{tRvsvarrho} then gives a general expression for the effective $d$-dimensional cosmological constant
\be
 V_{\rm eff} = \Bigl \langle \varrho  - \cZ  \Bigr \rangle_{\rm tot} \,.
\ee

\subsection{General static bulk solutions}
\label{subsec:Exactsolns}

This section presents the general bulk solutions for two branes. We start with the simple rugby-ball geometries that interpolate between two branes sourcing identical defect angles and then continue to the general case of two different branes. The solutions we find are all static -- actually maximally symmetric in the $d$ Lorentzian on-brane directions -- and symmetric under axial rotations about the two brane sources.

Rather than rewriting all of the field equations for the bulk region away from the branes, we note that these are easily obtained from the field equations of previous sections in the special case that $Z_\ssM = 0$ and $\psi = v$. Notice that $Z_\ssM = 0$ and $\psi = v$ already solve the $Z_\ssM$ and $\psi$ field equations, so it is only the others that need solutions, which must be the case since we have replaced the vortex degrees of freedom with an effective brane description.

These choices imply
\be
 L_{\rm kin} = L_\Psi = V_b = L_{\rm gm} = L_\ssZ = L_{\rm mix} = 0
 \quad\hbox{and so} \quad
 L_{\rm gge} = \check L_\ssA = L_\ssA = \frac12 \left( \frac{Q}{W^d} \right)^2 \,.
\ee
As a consequence of these we know
\be
 \cZ = 0 \,, \qquad
 \cX = \check \cX_\ssB = \Lambda - \check L_\ssA
 \qquad \hbox{and} \qquad
 \varrho = \check \varrho_\ssB = \Lambda + \check L_\ssA\,.
\ee

\subsubsection*{Rugby-ball geometries}
\label{section:Rugbyballgeom}

Because $\cZ = 0$ the solutions to the field equations can be (but need not be) locally maximally symmetric in the transverse 2 dimensions, rather than just axially symmetric there. For such solutions $W'$ must vanish and so the geometry is completely described by the constant scalar curvatures, $\check  R$ and $R$. The transverse dimensions are locally spherical, but with a defect angle at both poles corresponding to the removal of a wedge of the sphere.

Explicitly, we have $B = \alpha \ell \sin(\rho/\ell)$, and the polar defect angle has size $\delta = 2\pi(1 - \alpha)$. The sphere's curvature and volume are
\be
 R = \frac{2B''}{B} = - \frac{2}{\ell^2} \qquad \hbox{and} \qquad
 \cV_2 := 2\pi \int_0^{\pi\ell} \exd \rho \; B = 4 \pi \alpha \, \ell^2 \,,
\ee
where $\ell$ is the `radius' of the sphere. The relevant bulk field equations are the two Einstein equations
\be \label{Rdeq2}
 \check  R = - 2 \kappa^2 \left[ \Lambda - \frac12 \left( \frac{Q}{W^d} \right)^2 \right] \,,
\ee
and
\be \label{R2eq2}
  -R =  \frac{2}{\ell^2} =  2 \kappa^2 \left[ \frac{2\Lambda}{d} +  \left( 1 - \frac{1}{d} \right) \left( \frac{Q}{W^d} \right)^2 \right] \,,
\ee
with $Q$ fixed by flux quantization to be
\be \label{RBfluxQ}
 \frac{Q}{W^d} = \frac{\cN}{2 g_\ssA \alpha \, \ell^2}
  \qquad \hbox{where} \qquad
  \cN := N - n_{\rm tot} \varepsilon \left( \frac{g_\ssA}{e} \right) \,,
\ee
where $n_{\rm tot} = n_+ + n_-$ is the sum of the flux quantum for each vortex.\footnote{We take for simplicity the gauge coupling of the two vortices to be equal. See Appendix \ref{AppFluxQuantization} for a discussion of flux quantization for the $Z_\ssM$ and $A_\ssM$ fields.}

As shown in Appendix \ref{App:RugbyBalls}, the stable solution to these equations has compact transverse dimensions with radius
\be
 \frac{1}{\ell^2} = \frac{1}{r_\ssA^2} \left( 1 + \sqrt{ 1 - \frac{r_\ssA^2}{r_\Lambda^2} } \,\right)  \,,
\ee
where the two intrinsic length-scales of the problem are defined by
\be
 r^2_\Lambda := \frac{d}{4\kappa^2 \Lambda} \qquad \hbox{and} \qquad
 r_\ssA^2(\alpha) := \frac12 \left( 1 - \frac{1}{d} \right) \left( \frac{\cN \kappa}{g_\ssA \alpha} \right)^2 \,.
\ee
Clearly $\ell \simeq r_\ssA/\sqrt2$ when $r_\Lambda \gg r_\ssA$ and increases to $\ell = r_\ssA$ when $r_\Lambda = r_\ssA$. It is here that we first see why it is the combination $\cN \kappa/g_\ssA$ that sets the size of the extra dimensions. No solutions of the type we seek exist at all unless $r_\Lambda \ge r_\ssA$, which requires
\be \label{Lambdaxbound}
  \Lambda \le \frac{d-1}{2} \left( \frac{\alpha \, g_\ssA}{\cN \kappa^2} \right)^2 \,.
\ee
Finally, the on-brane curvature is
\bea \label{r0elltildeR}
 \check  R = - \frac{d}{2\, r_\Lambda^2} + \frac{d}{2(d-1)} \left( \frac{r_\ssA^2}{\ell^4} \right)
  &=& - \frac{d^2}{2(d-1)r_\Lambda^2}  + \left( \frac{d}{d-1} \right) \frac{1}{r_\ssA^2}  \left(1 + \sqrt{1 - \frac{r_\ssA^2}{r_\Lambda^2} } \right) \nn\\
  &=& \frac{d}{d-1} \left( - \frac{d}{2 r_\Lambda^2}  +  \frac{1}{\ell^2}  \right) \,,
\eea
which shows
\be
 \check  R \simeq \left(  \frac{2d}{d-1} \right) \frac{1}{r_\ssA^2} \qquad
 \hbox{has AdS sign when} \qquad r_\Lambda \gg r_\ssA \,,
\ee
but changes to dS sign
\be
 \check  R \to - \left( \frac{d-2}{d-1} \right)  \frac{2}{r_\ssA^2}  \qquad
 \hbox{when} \qquad r_\Lambda \to r_\ssA \,.
\ee
The on-brane curvature passes through zero when $\Lambda$ is adjusted to satisfy ${r_\ssA^2}/{r_\Lambda^2} =  {4(d-1)}/{d^2}$ (which is $\le 1$ for $d \ge 2$), and $\ell^2 = r_0^2 := (2/d) r_\Lambda^2$.

\subsubsection*{Geometries for general brane pairs}

Explicit closed-form solutions are also known where the branes at either end of the space have different properties. The difference between the two branes induces nontrivial warping and thereby breaks the maximal 2D symmetry of the transverse dimensions down to simple axial rotations.

The resulting solutions can be found by double Wick-rotating a charged black hole solution in $D$ dimensions \cite{SolnsByRotation, MSYK}, leading to the metric
\bea
 \exd s_0^2 &=& W^2(\wth) \, \check  g_{\mu\nu} \, \exd x^\mu \exd x^\nu + r_0^2\left(\frac{\exd \wth^2}{K(\wth)} + \wlmb^2 \, K(\wth) \sin^2\wth \,\exd \wvph^2\right) \nn\\
 &=& W^2(\wth) \,\check  g_{\mu\nu} \, \exd x^\mu \exd x^\nu  + r^2(\wth)\Big(\exd \wth^2 + \alpha^2(\wth) \, \sin^2\wth \,\exd \wvph^2\Big) \,,
\eea
where
\be
 W(\wth) := W_0 \Bigl( 1 + \eta\,\cos\wth \Bigr) \,, \qquad r(\wth) := \frac{r_0}{\sqrt{K(\wth)}} \quad \hbox{and} \quad \alpha(\wth) := \wlmb \, K(\wth) \,,
\ee
where $\eta$ is an integration constant and $r_0^{-2}: = 2\kappa^2\Lambda = (d/2)\, r_\Lambda^{-2}$. Notice that $r^2(\wth) \alpha(\wth) = r_0^2 \, \alpha_0$ for all $\wth$, and the vanishing of $g_{\theta\theta}$ implies the `radial' coordinate lies within the range $\vartheta_+ := 0 < \vartheta < \vartheta_- := \pi$. The geometry at the endpoints has a defect angle given by $\alpha_\pm = \alpha(\wth_\pm)$ and the derivative of the warp factor vanishes at both ends: $\exd W/\exd \vartheta \to 0$ as $\vartheta \to \vartheta_\pm$ (as required by the general Kasner arguments of earlier sections). In these coordinates the Maxwell field solves $\sqrt{-g} \; \wcF^{\wth\wvph} = Q$, which implies
\be
 \wcF_{\wth\wvph} = \frac{Q \, r_0^2 \, \alpha_0 \sin\wth}{W^d(\wth)}  \,.
\ee
Other properties of this metric --- including the explicit form for the function $K(\wth)$ --- are given in Appendix \ref{App:BeyondRugby}.

All told, the solution is characterized by three independent integration constants, in terms of which all other quantities (such as $\check  R$) can be computed. These constant can be taken to be $Q$ as well as an independent defect angle, $\alpha_+$ and $\alpha_-$, at each of the two poles. These three constants are themselves determined in terms of the source brane properties through the near-brane boundary conditions and the flux-quantization condition
\be \label{fqappwarp}
 \frac{\cN}{ g_\ssA \alpha_0 \, r_0^2} =  Q \int_0^\pi \!\exd\vartheta \,\frac{\sin\vartheta} {W^d(\vartheta)} \,,
\ee
where, as before, $\cN = N - n_{\rm tot}\varepsilon g_\ssA/e$ represents the vortex-modified flux-quantization integer.

\subsubsection*{Near rugby-ball limit}
\label{section:Nearrugby}

Although the general expressions are more cumbersome, it is possible to give simple formulae for physical quantities in terms of $Q$ and $\alpha_\pm$ in the special case where the geometry is not too different than a rugby ball. Because nonzero $\eta$ quantifies the deviation from a rugby-ball solution, in this regime we may expand in powers of $\eta$. In this section we quote explicit expressions that hold at low order in this expansion.

In the rugby-ball limit the functions $r(\wth)$, $\alpha(\wth)$ and $W(\wth)$ degenerate to constants, with $W(\wth) = W_0$ and $r(\wth) = \ell$ given explicitly in terms of $r_0^2 = (2/d) r_\Lambda^2$ and $\check  R$ by eq.~\pref{r0elltildeR}. Since $r(\wth)^2 \, \alpha(\wth) = r_0^2 \, \alpha_0$ this implies $\alpha_0$ is related to the limiting rugby-ball defect angle, $\alpha$, by
\be \label{ralpharbdefs}
  \alpha = \wlmb \left[ 1 + \left( \frac{d-1}{d} \right) \check  R \, r_0^2 \right] \,.
\ee

It happens that to linear order in $\eta$ the geometry near each pole takes the form
\be
 \exd s_{\pm}^2 \simeq W_0^2 (1 \pm 2 \,\eta) \exd s_4^2 + \ell_\pm^2 \Bigl[ \exd\wth^2 + \alpha_\pm^2 (\wth - \wth_\pm)^2 \, \exd\theta^2 \Bigr]
\ee
where
\bea
 \Delta W &:=& W_+ - W_- \simeq 2 \, W_0 \, \eta + \cO(\eta^2)\nn\\
 \ell_\pm^2  &\simeq&  \ell^2 \Bigl( 1 \pm \cC_\ssH \eta \Bigr) + \cO\left[\eta(\wth-\wth_\pm)^2,\eta^2 \right] \,,\quad \\
 \alpha_\pm &\simeq& \alpha \Bigl( 1 \mp \cC_\ssH \eta \Bigr) + \cO\left[\eta(\wth-\wth_\pm)^2,\eta^2 \right] \,.
\eea
with
\be
  \cC_\ssH := \frac{d - \frac23 + (d-1) \check  R \, r_0^2}{1-(d-1) \check  R \, r_0^2/d} \,.
\ee
This shows that the apparent rugby-ball radius and defect angle seen by a near-brane observer at each pole begins to differ for each brane at linear order in $\eta$.

To use these expressions to determine quantities in terms only of $Q$ and $\alpha_\pm$ requires knowledge of $\check  R$, and the field equations imply this is given for small $\eta$ by
\be \label{tRsmalleta}
 \check  R  = - \frac{1}{r_0^2} \left[ 1 + \left( \frac{2d-3}3 \right) \eta^2 \right] + \kappa^2 Q^2 \Bigl[ 1 + (d-1) \eta^2 \Bigr] + \cO(\eta^4) \,.
\ee
To complete the story we solve for $\eta$ in terms of $\alpha_\pm$ using
\be \label{etavsalpha}
 \frac{\alpha_- - \alpha_+}{\alpha}
 = 2\,\cC_\ssH \eta\,,
\ee
with $\alpha \simeq \frac12(\alpha_+ + \alpha_-)$, and use this to evaluate all other quantities.

For small $\eta$ the flux-quantization condition also simplifies, becoming
\be \label{fqappwarpsmeta}
 \frac{\cN}{2 g_\ssA \alpha_0 \, r_0^2} = \frac{\cN}{2 g_\ssA \alpha \, \ell^2} = \frac{Q}{2} \int_0^\pi \!\exd\vartheta \, \frac{\sin\vartheta}{W^d(\vartheta)} \simeq Q \left[ 1 + \frac{d(d+1)}6 \,\eta^2 + \cO(\eta^4) \right] \,.
\ee

\subsection{Relating bulk to vortex properties}
\label{subsec:bulkvortexmatching}

We see that the bulk solutions are determined by three parameters, $\alpha_\pm$ and $Q$. Earlier sections also show how these are related to the physical properties of the two source branes, since the near-brane defects are related to the renormalized brane tensions by
\be
 1 - \alpha_\pm = \frac{\kappa^2 \,W_\pm^d \check T_\pm}{2\pi} \,,
\ee
and $Q$ is determined in terms of brane properties by flux quantization, \pref{fqappwarp} (or, for small $\eta$, \pref{fqappwarpsmeta}).

\medskip\noindent{\em Parameter counting}

\medskip\noindent
An important question is to count parameters to see if there are enough integration constants in the bulk solutions to interpolate between arbitrary choices for the two vortex sources.

In total the source branes are characterized by a total of four physical choices: their tensions ({\em i.e.} defect angles) and localized flux quanta, $n_\pm$, to which we must add the overall flux quantum choice, $N$, for the bulk. But varying these only sweeps out a three-parameter set of bulk configurations because the flux choices ($n_\pm$ and $N$) only appear in the bulk geometry through flux quantization, and so only through the one combination, $\cN = N - \varepsilon (n_+ + n_-)(g_\ssA/e)$, that fixes $Q$. (Although they do not affect the geometry independent of $\cN$, the $n_\pm$ do govern the Bohm-Aharonov phase acquired by test particles that move about the source vortices.)

Consequently the three free constants --- $Q$ and $\alpha_\pm$ --- are sufficient to describe the static gravitational field set up by any pair of vortices, and once the brane properties (and $N$) are specified then all geometrical properties are completely fixed. The rugby ball geometries correspond to the special cases where $\check T_+ = \check T_-$.

This point of view, where the bulk dynamically relaxes in response to the presence of two brane sources, is complimentary to our earlier perspective which regarded integrating the field equations as an `evolution' in the radial direction (and so for which initial conditions at one brane completely specify the whole geometry --- and by so doing also fix the properties of the antipodal brane). They are equivalent because in the evolutionary point of view two of the integration constants to be chosen were $Q$ and $\check  R$, which are completely arbitrary from the perspective of any one brane. Their choices dictate the form of the interpolating geometry and correspond to the two-parameter family of branes (labeled by $\cN$ and $\alpha$) that could have been chosen to sit at the antipodal position.

\subsubsection*{On-brane curvature response}

Of particular interest is how the on-brane curvature, $\check  R$, responds to different choices for brane properties. In general this is given by \pref{tRvsvarrho}, in which we use the brane-vortex matching results --- \pref{rhobevalA} and \pref{XZbmatching} --- appropriate when the vortex size is negligibly small compared with the transverse KK scale,
$\cZ_b = \cX_b = 0$ and $\varrho_b  = W_b^d \, \check T_b \delta(y)/\sqrt{g_2}$, ensuring that
\be
 \langle \cZ \rangle_{\rm tot} \simeq 0 \,, \qquad
 \langle \cX \rangle_{\rm tot} \simeq \langle \Lambda - \check L_\ssA \rangle_{\rm tot} \qquad \hbox{and} \qquad
 \langle \varrho \rangle_{\rm tot} = \sum_b W_b^d \, \check T_b + \langle \Lambda + \check L_\ssA \rangle_{\rm tot} \,.
\ee
With these results \pref{tRvsvarrho} shows $\check  R$ takes the value that would be expected in $d$ dimensions in the presence of a cosmological constant of size $V_{\rm eff} = \langle \varrho \rangle_{\rm tot} = \left( 1 - 2/d \right) \langle \cX \rangle_{\rm tot}$, and so
\be \label{Veffbrane}
 V_{\rm eff} = \sum_b W_b^d \, \check T_b + \langle \Lambda + \check L_\ssA \rangle_{\rm tot} = \left( 1 - \frac{2}{d} \right) \langle \Lambda - \check L_\ssA \rangle_{\rm tot} \,,
\ee
In general $\check  R$ is not small. Since all quantities in $\varrho$ are positive (except perhaps for $\Lambda$), the resulting geometry is de Sitter-like unless cancelled by sufficiently negative $\Lambda$. Notice also that the second equality implies
\be
  \sum_b W_b^d \check T_b = - \frac{2}{d} \, \langle \Lambda \rangle_{\rm tot} - 2\left( 1 - \frac{1}{d} \right) \langle \check L_\ssA \rangle_{\rm tot} \,,
\ee
is always true. This states that for codimension-2 systems like this the `probe' approximation is never a good one: that is, it is {\em never} a good approximation to neglect the bulk response (the right-hand side) relative to the source tensions (the left-hand side) themselves.

\medskip\noindent{\em Near-flat response}

\medskip\noindent
Of particular interest are near-flat solutions for which $\Lambda$ is initially adjusted to cancel the rest of $\langle \varrho  \rangle_{\rm tot}$, after which brane properties are varied (without again readjusting $\Lambda$). One can ask how $\check  R$ responds to this variation.
To determine this response explicitly we use the near-rugby solution considered above, in the case where the unperturbed flat geometry is a rugby ball and for which the brane parameters are independently tweaked. To this end we take the initial unperturbed configuration to satisfy $W_0 = 1$ and
\be
 \Lambda = \Lambda_0:= \frac{Q_0^2}2 \quad \hbox{and} \quad
 \eta_0 =0 \quad\implies\quad \check  R_0 = 0
\ee
and then introduce small perturbations through $\delta \alpha$ and $\delta \cN$. From eq.~\pref{tRsmalleta}, we see immediately that
\be
 \delta \check  R = \frac2{r_0^2} \,\frac{\delta Q}{Q_0} + \cO(\eta^2)
\ee
and --- from the flux quantization condition in eq.~\pref{fqappwarpsmeta} --- we see that the leading perturbations are
\be
 \frac{\delta Q}{Q_0} = \frac{\delta \cN}{\cN_0} -\frac{\delta \alpha_0}{\alpha_0} + \cO(\eta^2) \,.
\ee
Lastly, since it is $\alpha = \frac12(\alpha_+ + \alpha_-)$ (and not $\alpha_0$) that is determined by $\check T_\pm$, we must use the relation --- eq.~\pref{ralpharbdefs} --- to write
\be
 \frac{\delta \alpha_0}{\alpha_0} = \frac{\delta\alpha}{\alpha} - r_0^2\left(1-\frac1d\right)\delta \check  R + \cO(\eta^2) \,.
\ee
Combining these formulae gives
\be
 \left(\frac{d-2}d\right)\delta \check  R \simeq \frac2{r_0^2} \left(\frac{\delta \alpha}{\alpha} - \frac{\delta \cN}{\cN}\right) \,,
\ee
to leading order, and so given perturbations of the form
\be
 \delta \alpha_\pm = -\frac{\kappa^2 \delta \check T_\pm}{2\pi} \,,\quad \delta \cN = - \frac{g_\ssA}{2\pi}\sum_{b\,\in \pm} \delta \xi_b \,,
\ee
the corresponding change in $\check  R$ has the form
\be
 \left( \frac{d-2}{d} \right) \delta \check  R \simeq -\frac{1}{r_0^2} \sum_{b = \pm} \left( \frac{\kappa^2 \delta \check T_b}{2\pi \alpha} - \frac{g_\ssA}{\pi\cN} \,\delta \xi_b \right) \,.
\ee
Comparing with \pref{tRvsVeff} --- and using $\kappa_d^{-2} = 4\pi \alpha r_0^2/\kappa^2$ and $\kappa^2 Q_0^2 r_0^2 = 1$ for the unperturbed flat rugby-ball geometry --- then shows that this curvature is what would have arisen from the $d$-dimensional vacuum energy
\be
  V_{\rm eff} = -\frac12\left(1-\frac 2d\right) \frac{\delta\check  R}{\kappa_d^2} \simeq \sum_{b = \pm}\left( \delta \check T_b - Q_0 \, \delta \xi_b\right) \,.
\ee
We see from this that when $\delta \cN = 0$ the curvature obtained is precisely what would be expected in $d$ dimensions given the energy change $\sum_b \check T_b$.

\section{Discussion}
\label{section:discussion}

In this paper we investigated the gravitational properties of branes that carry localized flux of a bulk field, or BLF branes. As noted in the introduction, the treatment of a gravitating BLF branes is not straightforward because the delta-like function used to represent their localization must depend on the proper distance away from the brane. Because of their particularly simple structure, this is not a problem for branes described only by their tension $\propto T$. However, the presence of metric factors in the BLF term $\propto \epsilon^{mn} A_{mn}$ complicates any calculation of transverse components of the brane's stress energy.

We resolved this ambiguity by constructing an explicit UV completion of BLF branes using Nielsen-Oleson vortices whose gauge sector mixes kinetically with a bulk gauge field. The gauge kinetic mixing, which is controlled by a dimensionless parameter $\varepsilon$, endows the bulk field with a non-zero flux in the localized region, even in the limit that this region is taken to be vanishingly small. This allows the UV theory to capture the effects of brane-localized flux.

The main result is that, in the UV picture, the gauge kinetic mixing can be diagonalized resulting in variables that clearly separate the localized sources from the bulk sources. In the diagonal basis, the energy associated with localized flux is always cancelled, and the canonical vortex gauge coupling is renormalized: $\hat e^2 = e^2 / (1 - \varepsilon^2)$. This allows us to identify the renormalized vortex tension as the quantity that controls the size of the defect angle in the geometry exterior to the vortex. We also find that the vortex relaxes to ensure that the average of the localized contributions to the transverse stress energy are controlled by the ratio between the size of the vortex and the characteristic bulk length scale $r_\varv / r_\ssB.$

This informs our treatment of the IR theory with branes. We find that the delta-function treatment of the brane is particularly useful for calculating the flux of the bulk field, including its localized contributions, and a delta-function shift in the bulk gauge field strength can diagonalize the brane-localized flux term. This change of variables endows the action with a divergent term that we can interpret as a renormalization of the brane tension, in analogy with the $e \to \hat e$ renormalization of the gauge coupling. We also show that the transverse components of the brane stress energy must vanish without explicitly calculating them. Rather, we use the Hamiltonian constraint and energy conservation to relate these stress energies to quantities which vanish as $r_\varv/r_\ssB \to 0$, thereby circumventing any ambiguity in the metric dependence of the corresponding brane interactions.

The techniques we employ here should be relevant to other brane couplings that contain metric factors. For example, there is a codimension-$k$ analogue of the BLF term that involves the Hodge dual of an $k$-form. Of particular interest is the case $k=1$ where the brane can couple to the derivative of a bulk scalar field $\phi$ as follows $  S_b \propto \int  \star \, \d \phi $, or a bulk gauge field $A$ as follows $S_b \propto \int \star A$. We have also provided an explicit regularization of a $\delta(0)$ divergence. These are commonplace in treatments of brane physics, and usually deemed problematic. However, there is likely a similar renormalization story in these other cases.

Lastly, we plan \cite{Companion} to also apply these techniques to a supersymmetric brane-world models that aim to tackle the cosmological constant problem \cite{SLED}.~The back-reaction of branes is a crucial ingredient of such models, and understanding the system in greater detail with an explicit UV completion will put these models on firmer ground and hopefully shed light on new angles from which to attack the CC problem.

\section*{Acknowledgements}

We thank Ana Acucharro, Asimina Arvanitaki, Savas Dimopoulos, Gregory Gabadadze, Ruth Gregory, Mark Hindmarsh, Stefan Hoffmann, Florian Niedermann, Leo van Nierop, Massimo Porrati, Fernando Quevedo, Robert Schneider and Itay Yavin for useful discussions about self-tuning and UV issues associated with brane-localized flux. The Abdus Salam International Centre for Theoretical Physics (ICTP), the Kavli Institute for Theoretical Physics (KITP), the Ludwig-Maximilian Universit\"at, Max-Planck Institute Garsching and the NYU Center for Cosmology and Particle Physics (CCPP) kindly supported and hosted various combinations of us while part of this work was done. This research was supported in part by funds from the Natural Sciences and Engineering Research Council (NSERC) of Canada, and by a postdoctoral fellowship from the National Science Foundation of Belgium (FWO), and by the Belgian Federal Science Policy Office through the Inter-University Attraction Pole P7/37, the European Science Foundation through the Holograv Network, and the COST Action MP1210 `The String Theory Universe'. Research at the Perimeter Institute is supported in part by the Government of Canada through Industry Canada, and by the Province of Ontario through the Ministry of Research and Information (MRI). Work at KITP was supported in part by the National Science Foundation under Grant No. NSF PHY11-25915.
\appendix

\section{Stress-energy conservation}
\label{appsec:SEConservation}

The matter field equations always guarantee the matter stress energy is covariantly conserved, $\nabla_\ssM T^{\ssM \ssN} = 0$. For the geometries of interest this has one nontrivial component, $\nabla_\ssM T^{\ssM \rho} = 0$, which implies
\be \label{SEconsrhoapp}
 \Bigl( B W^d \;{T^\rho}_\rho \Bigr)' = B W^d \left( \frac{B'}{B} \;  {T^\theta}_\theta + \frac{W'}{W} \; {T^\mu}_\mu \right) \,.
\ee
A useful way to rewrite this multiplies by $B$ and adds $BB'W^d {T^\rho}_\rho$ to both sides, so
\be \label{SEconsrho3}
 \Bigl( B^2 W^d \;{T^\rho}_\rho \Bigr)' = B W^d \left[ B' \; ( {T^\theta}_\theta + {T^\rho}_\rho ) + \frac{B W'}{W} \; {T^\mu}_\mu \right] \,,
\ee
or
\be \label{SEconsrho4}
 \Bigl[ \sqrt{-g} \; B (\cZ - \cX) \Bigr]' = - \sqrt{-g} \left[ 2B' \cX + \frac{d B W'}{W} \; \varrho \right] \,.
\ee
When applied to a vortex on flat space --- for which $W = B' = 1$ and the constraint \pref{constraint} implies $\cZ - \cX = 0$ outside the vortex --- integrating eq.~\pref{SEconsrho4} over the vortex reduces to the simple statement
\be
 \left. \bigl \langle \cX \bigr \rangle_\varv \right|_{\rm flat} = \left. \bigl \langle L_{\rm pot} - L_{\rm gge} \bigr \rangle_\varv \right|_{\rm flat}  = 0 \,,
\ee
a result that may also be derived as the vortex equation of motion corresponding to extremizing the flat-space action against rigid rescalings.

\section{Flux quantization}
\label{AppFluxQuantization}

For compact transverse dimensions the underlying transverse geometry of interest has the topology of a sphere, leading to two flux-quantization conditions; one for each of the $U(1)$ gauge fields.

Our interest is in vortices that are much smaller than the size of the transverse space. In this case we take the complex vortex field and the gauge field $Z_\ssM$  at the equator to be gauge-equivalent to their vacuum values, $\Psi = v e^{i e u}$ and $Z_\ssM = \partial_\ssM u$. Here single-valuedness of $\Psi_\pm$ on both patches, $\cP_\pm$, implies
\be
 u_\pm(\theta + 2\pi) = u_\pm(\theta) + \frac{2\pi n_\pm}{e} \,,
\ee
for some integers $n_\pm$. The choices for $n_\pm$ can be chosen differently because they differ by a gauge transformation, $g = e^{ie\omega}$, whose single-valuedness implies $\omega(\theta + 2\pi) = \omega(\theta)+ 2\pi N_\ssZ$, provided $N_\ssZ = n_+ + n_- =: n_{\rm tot}$.

The total $Z$-flux through each hemisphere is related to the integral of $Z_\theta$ around the equator by
\be
 \Phi_{\ssZ\pm} = \int_{\cP_\pm} \exd Z  = \oint_{\partial\cP_\pm} Z = \frac{2\pi n_\pm}{e} \,,
\ee
and so the total $Z$-flux through the sphere is
\be
 \Phi_{\ssZ} = \frac{2\pi n_{\rm tot}}{e}  \,.
\ee

For the $A_\ssM$ gauge field we imagine test charges situated far from the vortices that couple to $A_\ssM$ and carry charge $g_\ssA$. The action for such a charge probe is
\be
 S_{\rm probe} = g_\ssA \int A \,,
\ee
where the integration is along the world-line of the charge. Moving such a charge around the equator far from the vortex contributes an amount
\be
 \exp \left[ i g_\ssA \oint_{\rm eq} A \right] = \exp \left[ i g_\ssA \int_{\cP_+} \exd A \right] = \exp \left[ -i g_\ssA \int_{\cP_-} \exd A \right] \,,
\ee
to the path integral, where the two equalities rewrite the integral using Stoke's theorem and the observation that the equator can be considered to be the boundary of either hemisphere (with the sign keeping track of the orientation of the boundary). In order for this phase to be single-valued in the path integral we must therefore demand the fluxes, $\Phi_{\ssA\pm} = \int_{\cP_\pm} \exd A$, satisfy
\be
 \Phi_\ssA := \Phi_{\ssA+} + \Phi_{\ssA-} = \frac{2\pi N}{g_\ssA} \,,
\ee
for some integer $N$.

\subsubsection*{Bulk vs localized flux}

Suppose now we take a test charge that starts life coupled only to $A_\ssM$ and move it around the vortex, keeping always far enough from the vortex that the $Z_\ssM$ magnetic field is negligible. Then we define the flux seen by this charge by
\bea \label{AfluxV2}
 \Phi_\ssA(X_\varv) &:=& \oint\limits_{\partial X_\varv} A = \int\limits_{X_\varv} \exd A = \int\limits_{X_\varv} \Bigl( \exd \check A - \varepsilon  \,\exd Z \Bigr) \nn\\
 &=& 2\pi \left[ \int_{0}^{\rho_\varv} \exd \rho \left( \frac{Q B}{W^d} \right)  +\frac{n \, \varepsilon}{e} \right]
 \simeq \frac{2\pi n}{e} \,,
\eea
where the second-last equality uses flux quantization (for integer $n$) for the vortex solution for $Z$ localized well within region $V$:
\be \label{ZfluxV}
 \Phi_\ssZ(X_\varv) := \int\limits_{X_\varv} \exd Z = -\frac{2\pi n}{e} \,,
\ee
and the sign on the far right-hand side is chosen for later convenience. The approximate equality in \pref{AfluxV2} drops the order $(\rho_\ssV/\ell)^2$ contribution of the $\check A$ flux over the vortex volume relative to the localized $Z$ flux.

What is important about \pref{AfluxV2} is that the gauge-field mixing implies that the test charge now sees a vortex-localized component due to the appearance of the $Z$ term. It is in this sense that our system provides a UV completion for branes carrying nonzero brane-localized flux.

On general grounds the flux of $A$ is also quantized, and this fixes the value of $Q$. That is, if the integration is performed over the entire transverse dimensions it must satisfy
\be
 \Phi_\ssA(\hbox{tot}) = \frac{2\pi N}{g_\ssA} \,,
\ee
where $N$ is an integer and $g_\ssA$ is the gauge coupling of the field $A_\ssM$ to its test charge. Consequently, the presence of the brane-localized flux modifies what flux quantization demands for $Q$:
\be
     Q \int_{0}^{\ell_\star} \exd \rho \left( \frac{B}{W^d} \right) = \frac{N}{g_\ssA} - \frac{n_{\rm tot} \, \varepsilon}{e}  \,.
\ee
where $\rho = \ell_\star$ denotes the proper distance between the branes (defined by the two places where $B$ vanishes: $B(0) = B(\ell_\star) = 0$), and $n_{\rm tot} = n_+ + n_-$ is the sum of the flux quanta for the vortices localized at each of these positions.

\section{Solutions}
\label{SolutionsApp}

This appendix describes more details about the approximate and exact solutions described in the main text.

\subsection{Approximate near-vortex solutions}
\label{KasnerApp}

For the purposes of matching the bulk integration constants to the vortex properties we are most interested in the form of the solutions very near to, but outside of, the vortex sources. We start by recapping the form of the bulk solutions very close, but outside of, a small vortex.

\subsubsection*{Asymptotic forms}

Near the branes it is possible to expand the solutions in powers of $\rho/r_\ssB$, where $\rho$ denotes proper distance in the bulk geometry from the vortex. Writing, as before, the metric in the form
\be
 \exd s^2 = W^2(\rho) \, \check  g_{\mu\nu} \, \exd x^\mu \exd x^\nu + \exd \rho^2 + B^2(\rho) \, \exd \theta^2 \,,
\ee
we seek near-vortex solutions to the Einstein equations of the form
\bea \label{powerformsapp}
 W &=& W_0 \left( \frac{\rho}{r_\ssB} \right)^w + W_1 \left( \frac{\rho}{r_\ssB} \right)^{w+1} + W_2 \left( \frac{\rho}{r_\ssB} \right)^{w+2} + \cdots  \,, \nn\\
 B &=& B_0 \left( \frac{\rho}{r_\ssB} \right)^b  + B_1 \left( \frac{\rho}{r_\ssB} \right)^{b+1} + B_2 \left( \frac{\rho}{r_\ssB} \right)^{b+2} + \cdots  \,,
\eea
and so on. The special case of flat space in polar coordinates corresponds to $w = 0$ and $b = 1$, without the need for higher powers of $\rho/r_\ssB$.

The leading powers, $w$ and $b$, are constrained by the leading terms in the expansion of the field equations around the vortex position, $\rho = 0$. The source terms on the RHS of the Einstein equations in the bulk involve $\Lambda$ and $\check L_\ssA = \frac12 (Q/W^d)^2$, which vary respectively like $\rho^0$ and $(\rho/r_\ssB)^{-2dw}$ as $\rho \to 0$. By comparison, as $\rho \to 0$ the curvature on the LHS of the Einstein equation are
\bea
 \cR_{(d)} - \frac{\check  R}{W^2} &=&  d \left[ (d-1) \left( \frac{W'}{W} \right)^2 + \left( \frac{W''}{W} + \frac{B'W'}{BW} \right) \right]  \nn\\
 &=&  d \left\{ (d-1) \left( \frac{w}{\rho} \right)^2 + \left[ \frac{w(w-1)}{\rho^2} + \frac{bw}{\rho^2} \right] \right\} \left[ 1 + \cO \left( \frac{\rho}{r_\ssB} \right) \right]\nn\\
 &=&  dw \left( \frac{dw + b - 1}{\rho^2} \right) \left[ 1 + \cO \left( \frac{\rho}{r_\ssB} \right) \right] \,.
\eea
Assuming $w < 1$ --- so that $\check  R/W^2 \propto (\rho/r_\ssB)^{-2w}$ is subdominant to the $1/\rho^2$ term explicitly displayed (a result justified below) --- we see that the ($\mu\nu$) Einstein equation implies $w(dw+b-1) = 0$. Similarly,
\be
 {\cR^\theta}_\theta = \frac{B''}{B} + d \left( \frac{B'W'}{BW} \right)
   = \frac{b(dw + b - 1)}{\rho^2} \left[ 1 + \cO \left( \frac{\rho}{r_\ssB} \right) \right] \,,
\ee
implies $b(dw+b-1) = 0$, and
\be
 {\cR^\rho}_\rho = \frac{B''}{B} + d \left( \frac{W''}{W} \right)
   = \frac{b(b-1) + dw(w-1)}{\rho^2} \left[ 1 + \cO \left( \frac{\rho}{r_\ssB} \right) \right] \,,
\ee

Besides the trivial special case ($w=b=0$) we see that the vanishing of the $1/\rho^2$ terms in the field equations implies the following two Kasner conditions:
\be \label{Kasnerlinapp}
 dw + b = 1 \,,
\ee
and
\be \label{Kasnerquadapp}
 d w^2 + b^2 = 1 \,.
\ee
The last of these in turn implies $w$ and $b$ must reside within the intervals
\be \label{limitsapp}
 |w| \le \frac{1}{\sqrt{d}} \qquad \hbox{and} \qquad
 |b| \le 1 \,,
\ee
which shows in particular why $1/W^2 \propto (\rho/r_\ssB)^{-2w}$ is less singular than $1/\rho^2$, as assumed above. The Kasner conditions, eqs.~\pref{Kasnerlinapp} and \pref{Kasnerquadapp} have precisely two solutions: either $w = 0$ and $b = 1$ (as is true for the rugby-ball solutions described above) or $dw = 1$ and $b = 0$.

\subsection{Exact Solutions}
\label{App:Solutions}

This section explores some properties of the solutions to the vortex-bulk field equations.

\subsubsection*{Rugby-ball geometries}
\label{App:RugbyBalls}

We next describe some details associated with the rugby-ball geometries, for which solutions to the field equations with $\psi = v$ and $Z_\ssM = 0$ are sought of the form $B = \alpha \ell \sin(\rho/\ell)$, with $W$ constant. The transverse curvature and volume are
\be
 R = \frac{2B''}{B} = - \frac{2}{\ell^2} \qquad \hbox{and} \qquad
 \cV_2 := 2\pi \int_0^{\pi\ell} \exd \rho \; B = 4 \pi \alpha \, \ell^2 \,,
\ee
where $\ell$ is the `radius' of the sphere. The relevant bulk field equations are the two Einstein equations
\be \label{Rdeq2app}
 \check  R = - 2 \kappa^2 \left[ \Lambda - \frac12 \left( \frac{Q}{W^d} \right)^2 \right] \,,
\ee
and
\be \label{R2eq2app}
  -R =  \frac{2}{\ell^2} =  2 \kappa^2 \left[ \frac{2\Lambda}{d} +  \left( 1 - \frac{1}{d} \right) \left( \frac{Q}{W^d} \right)^2 \right] \,,
\ee
with $Q$ fixed by flux quantization to be
\be \label{RBfluxQapp}
 \frac{Q}{W^d} = \frac{\cN}{2 g_\ssA \alpha \, \ell^2}
  \qquad \hbox{where} \qquad
  \cN := N - n_{\rm tot} \varepsilon \left( \frac{g_\ssA}{e} \right) \,.
\ee

Using \pref{RBfluxQapp} in \pref{R2eq2} allows $\ell^2$ to be solved as a function of brane properties (which enter through the defect-angle parameter $\alpha$). Defining
\be
 r^2_\Lambda := \frac{d}{4\kappa^2 \Lambda} \qquad \hbox{and} \qquad
 r_\ssA^2(\alpha) := \frac12 \left( 1 - \frac{1}{d} \right) \left( \frac{\cN \kappa}{g_\ssA \alpha} \right)^2 \,,
\ee
we find
\be
 \frac{r_\ssA^2}{\ell^4} - \frac{2}{\ell^2} + \frac{1}{r_\Lambda^2} = 0 \qquad\hbox{or, equivalently} \qquad \ell^4 - 2 r_\Lambda^2 \ell^2 + r_\Lambda^2 r_\ssA^2 = 0 \,,
\ee
which has solutions
\be
 \frac{1}{\ell^2_\pm} = \frac{1}{r_\ssA^2} \left( 1 \pm \sqrt{ 1 - \frac{r_\ssA^2}{r_\Lambda^2} } \,\right) \qquad\hbox{or, equivalently} \qquad \ell^2_\pm = r_\Lambda^2 \left( 1 \mp \sqrt{1 - \frac{r_\ssA^2}{r_\Lambda^2}} \,\right)
 \,.
\ee
Clearly $\ell_- \simeq \sqrt2 \; r_\Lambda$ and $\ell_+ \simeq r_\ssA/\sqrt2$ when $r_\Lambda \gg r_\ssA$. As $r_\Lambda$ decreases from infinity $\ell_-$ also decreases and $\ell_+$ increases until they meet at $\ell_+ = \ell_- = r_\ssA = r_\Lambda$ when $r_\Lambda = r_\ssA$. No solutions of the type we seek exist if $r_\Lambda < r_\ssA$, and so the existence of solutions requires we choose
\be \label{Lambdaxboundapp}
  \Lambda \le \frac{d-1}{2} \left( \frac{\alpha \, g_\ssA}{\cN \kappa^2} \right)^2 \,.
\ee
Finally, the corresponding on-brane curvature becomes
\be
 \check  R_\pm = - \frac{d}{2\, r_\Lambda^2} + \frac{d}{2(d-1)} \left( \frac{r_\ssA^2}{\ell_\pm^4} \right)
  = - \frac{d^2}{2(d-1)r_\Lambda^2}  + \left( \frac{d}{d-1} \right) \frac{1}{r_\ssA^2}  \left(1 \pm \sqrt{1 - \frac{r_\ssA^2}{r_\Lambda^2} } \right) \,,
\ee
which shows
\be
 \check  R_- \simeq - \left[ \frac{d(2d-1)}{2(d-1)} \right] \frac{1}{r_\Lambda^2} \,, \quad \hbox{and} \quad
 \check  R_+ \simeq \left(  \frac{2d}{d-1} \right) \frac{1}{r_\ssA^2} \qquad
 \hbox{if} \quad r_\Lambda \gg r_\ssA \,.
\ee
These two roots converge towards one another as $r_\Lambda \to r_\ssA$, eventually converging at
\be
 \check  R = - \left( \frac{d-2}{d-1} \right)  \frac{2}{r_\ssA^2}  \qquad
 \hbox{when} \qquad r_\Lambda = r_\ssA \,.
\ee
Although $\check  R_-$ remains negative (de Sitter space) between these two limits, $\check  R_+$ changes sign. It passes through zero when
\be
 \frac{r_\ssA^2}{r_\Lambda^2} =  \frac{4(d-1)}{d^2} \qquad (\hbox{which is } \le 1 \hbox{ for } d \ge 2 )\,.
\ee

The origin of the two roots can be understood by thinking of the lagrangian as a function of $\ell$ in order to understand the potential that $\ell$ is minimizing. Since $L_{\rm tot} = L_\EH + \check L_\ssA + \Lambda$, and since $\cR_{(2)} = R = -2/\ell^2$ and $\check L_\ssA \propto r_\ssA^2/\ell^4$, and since $\sqrt{g_2} \propto \ell^2$, we can see that in the 4D Einstein frame we have a potential for $\ell$ that involves three terms:
\be
 V_{\rm eff} (\ell) = \frac{a}{r_\Lambda^2 \ell^2} - \frac{b}{\ell^4} + \frac{c \, r_\ssA^2}{\ell^6} \,,
\ee
where $a$, $b$ and $c$ are positive dimensionless constants (where positive $a$ assumes positive $\Lambda$) and an overall factor of $1/\ell^2$ comes from the transition to 4D Einstein frame. This implies the potential climbs to positive values as $\ell$ comes in from infinity, eventually reaching a maximum and then falling to a minimum before climbing again to infinity as $\ell \to 0$.

We can see that the solution $\ell_-$ and $\check  R_-$ describe the {\em maximum} of this potential in the low-energy theory, and this is why it always occurs at positive values of $V_{\rm eff}$. It should therefore be unstable. It is the solution $1/\ell_+$ and $\check  R_+$ that describes the minimum, and whether this occurs for positive or negative values of $V_{\rm eff}$ depends on the detailed size of the parameters. The condition \pref{Lambdaxbound} is the condition for the existence of both a local maximum and minimum, and when it is not satisfied the potential simply rises monotonically as $\ell$ falls. (We also see that a minimum exists, but always with negative potential, even if $\Lambda \to 0$ provided $r_\ssA \ne 0$.)

\subsubsection*{Beyond Rugby Balls}
\label{App:BeyondRugby}

We here record the properties of the more general bulk solutions appropriate when the source branes are not identical. Although the solutions are described constructively here, they may also be found by double Wick-rotating the higher-dimensional black-hole solution \cite{SolnsByRotation}.

We start with our standard ansatz, with radial coordinate chosen so that $B \propto 1/A = F(\wc)$; that is,
\be
 \exd s^2 = W^2(\wc) \, \check  g_{\mu\nu}
 \, \exd x^\mu \exd x^\nu + \frac{\exd \wc^2}{F(\wc)} + F(\wc) \, c_0^2\,\exd \wvph^2 \,.
\ee
In these coordinates the ($\wc\wc$) -- ($\theta\theta$) Einstein equation, together with $\cZ = 0$ in the bulk, becomes
\be
 \frac{1}{B} \left( \frac{\exd W}{\exd \rho} \right) =
 \frac{1}{AB} \left( \frac{\exd W}{\exd \wc} \right) = k \,,
\ee
for constant $k$, and so because $AB$ is $\wc$-independent we must have $W(\wc) \propto \wc$. This leaves
\be
 \exd s^2 = \left(\frac{\wc}{\wc_0}\right)^2 \check  g_{\mu\nu}
 \, \exd x^\mu \exd x^\nu + \frac{\exd \wc^2}{F(\wc)} + F(\wc) \, c_0^2\,\exd \wvph^2 \,,
\ee
where $\wc_0$ --- or, equivalently, $k$ --- can be absorbed into a rescaling of the coordinate $\wc$.

The Maxwell field is also given in terms of these metric functions because the Maxwell equation is solved by $\sqrt{-g} \; \wcF^{\wc\wvph} = Q$, which implies
\be
 \wcF_{\wc\wvph} = Q\,c_0\left(\frac{\wc_0}\wc\right)^d \,.
\ee
For definiteness we take the maximally symmetric $d$-dimensional subspace to be de Sitter (or flat) space with Hubble scale $H$, so $\check  R_{\mu\nu} = - (d-1) H^2\, \check  g_{\mu\nu}$ and
\be
 \cR_{(d)} = -\frac{d(d-1)\,H^2}{(\wc/\wc_0)^2} + d \left( d-1 \right) \frac{F(\wc)}{\wc^2} + \frac {d}{\wc}\,\pd_\wc F \,.
\ee
This is to be used in the Einstein equation $\cR_{(d)} = -2\kappa^2 \cX$, with
\be
 \cX = \Lambda - \frac12 \left( \frac{Q}{W^d} \right)^2 = \Lambda - \frac{\wQ^2}{2} \left(\frac{\wc_0}\wc\right)^{2d} \,,
\ee
which implies $F(\wc)$ must take the form
\be
 \frac{F(\wc)}{\wc_0^2} = H^2 -A\,\left(\frac\wc{\wc_0}\right)^2
 + \left(\frac{\wc_0}\wc\right)^{d-1}B - \left(\frac{\wc_0}\wc\right)^{2(d-1)}C
\ee
where
\be
 A:= \frac{2\,\kappa^2\Lambda}{d(d+1)} \,,\quad C:= %%\frac1{d(d-1)}\,\frac{\kappa^2\wQ^2}{g^2} \,.(assume g is gauge coupling, and you are normalizing the kinetic term differently?)
 \frac{\kappa^2\wQ^2}{d(d-1)}\, \,,
\ee
and $B$ is an integration constant.

The constants $B$ and $H$ can be traded for two other parameters, $\wc_+$ and $\wc_-$, that define the zeroes of $F$: {\em i.e.} $F(\wc_\pm)=0$. Extracting two factors that enforce this vanishing, allows one to define
\be
 F(\wc) = A (\wc_+-\wc)(\wc-\wc_-) \,G(\wc) \quad{\rm where}\quad G(\wc) = 1+ \sum_{n=1}^{2(d-1)} \frac{G_n}{\wc^n}
\ee
with
\be
 G_n = \left\{\begin{matrix}\displaystyle
 \bar\wc_n \,,& n=1\\
 \bar\wc_n - \frac{H^2}A\wc_0^2\,\bar\wc_{n-2} \,,& n=2,\ldots, d \\
 \bar\wc_n - \frac{H^2}A\wc_0^2\,\bar\wc_{n-2} - \frac{B}A\,\wc_0^{d+1} \,\bar\wc_{n-d-1}\,, & n=d+1,\ldots, 2(d-1)
 \end{matrix}\right.
\ee
given
\be
 \bar\wc_n := \sum_{i=0}^n \wc_+^i \wc_-^{n-i} \,.
\ee
$H$ and $B$ are given explicitly in terms of $\wc_\pm$ by
\begin{align}
 H^2 &= \left[ \frac{\left({\wc_+}/{\wc_0}\right)^{d+1} - \left( {\wc_-}/{\wc_0}\right)^{d+1}}{\left( {\wc_+}/{\wc_0}\right)^{d-1} - \left({\wc_-}/{\wc_0}\right)^{d-1}} \right] A - \left(\frac{\wc_+}{\wc_0}\right)^{1-d} \left(\frac{\wc_-}{\wc_0}\right)^{1-d}C \\
 \hbox{and} \quad
 B &= \left[ \frac{ \left( {\wc_+}/{\wc_0} \right)^{2} - \left( {\wc_-}/{\wc_0} \right)^{2} }{ \left( {\wc_+}/{\wc_0} \right)^{1-d} - \left( {\wc_-}/{\wc_0} \right)^{1-d}} \right] A + \left[ \left( \frac{\wc_+}{\wc_0} \right)^{1-d} + \left( \frac{\wc_-}{\wc_0} \right)^{1-d} \right] C \,.
\end{align}
Since $H^2\geq0$, the above imply
\be
 B \geq \left[ \frac{ \left( {\wc_+}/{\wc_0} \right)^{2d} - \left( {\wc_-}/{\wc_0} \right)^{2d} }{ \left( {\wc_+}/{\wc_0} \right)^{d-1} - \left( {\wc_-}/{\wc_0} \right)^{d-1}} \right] C > 0 \quad \forall \quad \wc_+>\wc_- \;(\mathrm{whenever}\; d>1) \,.
\ee
As a check, consider the limiting case where $\wc_\pm,\wc_0 \to 1$ and $H\to 0$. Then, since $\bar\wc_n \to n+1$ in that limit, we find
\be
 C=\left(\frac{d+1}{d-1}\right)A \,,\quad B=\left(\frac{2d}{d-1}\right)A\,,\quad G_n = \left\{\begin{matrix}\displaystyle
 2 \,, & n=1\\
 (n+1)  \,,& n=2,\ldots, d \\
 (n+1)  - \frac{B}A(n-d)\,, & n=d+1,\ldots ,2(d-1)
 \end{matrix}\right.
\ee
and so
\be
G(\wc)\to  d(d+1) -(d-1)^2 \frac{H^2}A\,.
\ee
Therefore --- in this limit --- the metric becomes that of a rugby ball with radius, $r$, and defect parameter, $\alpha$, given by
\be
r^2 = \frac{r_0^2}{1-[(d-1)Hr_0]^2} \,,\quad \alpha = \alpha_0 \big(1-[(d-1)Hr_0]^2\big) \,,
\ee
where $r_0^{-2} := d(d+1) A = 2\kappa^2 \Lambda$, which is independent of $d$.

From here, we wish to change variables from $\wc$ to $\wth$ such that $\wc=\wc_\pm$ is identified with $\cos\wth = \pm 1$; we find that the transformation
\be
 \frac{\wc}{\wc_0} =  W(\wth) \,,\quad W(\wth) :=1+\eta\,\cos\wth
\ee
where
\be
 \wc_0 = \frac{\wc_++\wc_-}2\quad{\rm and}\quad \eta = \frac{\wc_+-\wc_-}{\wc_++\wc_-}
%A= \frac{\wc_++\wc_-}{2\bar \wc_0} \,,\quad B = \frac{\wc_--\wc_+}{2\wc_0} \,,\quad \wc_0 := \sqrt{\wc_+\wc_-}
\ee
fills the bill.
%(The double appearance of $\wc_0$ --- both in the above and in the original metric ans\"atz --- is intentional; it is done to cancel the previous one in the result for the metric.)
%(The normalization of $A$ and $B$ has been chosen such that $A^2-B^2=1$.)
Under this coordinate change, we see that
\be
 \exd\wc^2 = (\wc_+-\wc)(\wc-\wc_-)\,\exd\wth^2 = \left(\frac{\wc_+-\wc_-}2\right)^2 \sin^2\wth \,\exd\wth^2
\ee
and so the metric becomes
\begin{align}
 \exd s_0^2 &= W^2(\wth) \, \check  g_{\mu\nu} \, \exd x^\mu \exd x^\nu + r_0^2\left(\frac{\exd \wth^2}{K(\wth)} + K(\wth)\wlmb^2 \sin^2\wth \,\exd \wvph^2\right) \\
 &= W^2(\wth) \,\check  g_{\mu\nu} \, \exd x^\mu \exd x^\nu  + r^2(\wth)\Big(\exd \wth^2 + K^2(\wth)\wlmb^2 \sin^2\wth \,\exd \wvph^2\Big)
\end{align}
where we identify $r_0^{-2}:=d(d+1)A=2\kappa^2\Lambda$ and $r(\wth)=r_0/K^{1/2}(\wth)$, and where $K:=\frac{G}{d(d+1)}$. Furthermore, since the expressions for $H^2$, $B$, and $\bar\wc_n$ differ from their unwarped values only quadratically in $\eta$, {\rm i.e.}
\begin{align}
H^2 &=\left(\frac{d+1}{d-1}\right)\left(1+\frac{(2d-3)}3\,\eta^2\right)A-\left[1+(d-1)\eta^2\right]C+\cO(\eta^4) \\
B &= \frac 2{(1-d)} \left(1-\frac{d(d+1)}6\,\eta^2\right)A + \left[2+d(d-1)\eta^2\right]C+\cO(\eta^4) \\
\bar\wc_n &= (n+1)\left(1+\frac{n(n-1)}6\,\eta^2+\cO(\eta^4)\right)\,,
\end{align}
the leading-order corrections to $K(\wth)$ arise from the $\eta$-dependence of the warp factors:
\begin{align}
K(\wth) &\simeq 1-[(d-1)Hr_0]^2 -\frac1{d(d+1)}\bigg(\sum_{n=1}^{2(d-1)}n\,G_{n0}\bigg)\eta\,\cos\wth + \cO(\eta^2)  \\
&\simeq 1-[(d-1)Hr_0]^2 -\left(d-\frac23 - d[(d-1)Hr_0]^2\right)\eta\,\cos\wth + \cO(\eta^2) \,.
\end{align}
This means that --- if $\eta\ll 1$ --- then $K_\pm:=K(\wth_\pm)$ is well-approximated by
\begin{align}
K_\pm &\simeq 1\mp\left(d-\frac23\right)\,\eta -[(d-1)Hr_0]^2(1\mp d\,\eta)+\cO(\eta^2) \\
&= \Big(1-[(d-1)Hr_0]^2\Big)\bigg[1 \mp\left(d-\frac23\right)\underbrace{\left(\frac{1-\tfrac d{d-2/3}[(d-1)Hr_0]^2}{1-[(d-1)Hr_0]^2}\right)\eta}_{:=\eta_\ssH} \bigg] + \cO(\varepsilon^2) \,.
\end{align}
Then, since the metric takes the form
\be
\exd s_{\pm}^2 \simeq (1\pm2\,\eta)\exd s_4^2 + r_\pm^2\Big(\exd\wth^2 + \alpha_\pm^2 (\wth-\wth_\pm)^2\,\exd\theta^2\Big)
\ee
near each pole (to linear order in $\varepsilon$), we find
\begin{align}
r_\pm &= \frac{r_0}{K_\pm^{1/2}} \simeq  r \left[1\pm\left(\frac d2 - \frac13\right)\eta_\ssH\right]+\cO[\eta(\wth-\wth_\pm)^2,\eta^2] \,,\quad \\
\alpha_\pm &= \wlmb K_\pm \simeq \alpha\left[1\mp\left(d-\frac23\right)\eta_\ssH\right]+\cO[\eta(\wth-\wth_\pm)^2,\eta^2] \,.
\end{align}
%
%Therefore, for small warping, there is a cross-over point where the influence of warping (in increasing/decreasing the local radius and defect angle) reverses its polarity; this occurs at
%%
%
%%
%We also see that --- at this critical value --- the effects of 4D de Sitter expansion and warping exactly cancel in their effect on the bulk geometry near the poles. That is, the geometry of a system with critical $H$ and some perturbatively-small warping can masquerade as a rugby ball near each defect.

Rearranging the expression for $\alpha_\pm$ and identifying $\kappa^2 L_\pm/(2\pi) = 1-\alpha_\pm$, we see that
\be
\alpha \simeq \frac{\alpha_++\alpha_-}2 = 1-\frac{\kappa^2}{4\pi}(L_++L_-) + \cO[(\kappa^2 L_\pm)^2] \,,\quad \eta_\ssH \simeq \frac{\kappa^2/4\pi}{(d-2/3)}(L_+-L_-) + \cO[(\kappa^2 L_\pm)^2] \,.
\ee
Furthermore, we see that the warp factor changes by an amount
\be
\Delta W := W(\pi) - W(0) = 2\varepsilon =\frac{\kappa^2/2\pi}{(d-2/3)}(T_+-T_-)\left(\frac{1-[(d-1)Hr_0]^2}{1-\tfrac d{d-2/3}[(d-1)Hr_0]^2}\right)
\ee
across the bulk, for $H < H_{\rm crit}$ where
\be \label{Hcrit}
[(d-1)H_{\rm crit}r_0]^2 = 1-\frac2{3d} \quad\leftrightarrow\quad \check  R_{\rm crit} = -\left(\frac{d-2/3}{d-1}\right)2\kappa^2\Lambda \,.
\ee

\end{document}